\documentclass[lettersize,journal]{IEEEtran}
\usepackage{amsmath,amssymb,amsfonts,amsthm}
\usepackage{algorithmic}
\usepackage{algorithm}
\usepackage{array}
\usepackage{textcomp}
\usepackage{stfloats}
\usepackage{url}
\usepackage{verbatim}
\usepackage{graphicx}
\usepackage{mathtools,amsmath,amsthm,amssymb,amsfonts}
\usepackage[backend=biber,style=ieee,sortlocale=en-US,url=false,doi=false,date=year,minnames=1,maxnames=5,isbn=false, sortcites, dashed=false]{biblatex}
\usepackage[binary-units=true]{siunitx}
\usepackage{graphicx}
\usepackage{booktabs}
\usepackage{caption}
\usepackage{subcaption}
\usepackage{textcomp}
\usepackage{csquotes}
\usepackage{multirow}
\usepackage{hyperref}
\usepackage{siunitx}

\newtheorem{example}{Example}

\newtheorem{definition}{Definition}

\begin{document}

\title{Shuttling for\\Scalable Trapped-Ion Quantum Computers}

\author{Daniel Schoenberger,~\IEEEmembership{Student Member,~IEEE,} Stefan Hillmich,~\IEEEmembership{Member,~IEEE,} Matthias Brandl, \\and Robert Wille,~\IEEEmembership{Senior Member,~IEEE}}

\maketitle

\hypersetup{
	pdftitle={Shuttling for Scalable Trapped-Ion Quantum Computers},
	pdfsubject={},
    pdfauthor={Daniel Schoenberger, Stefan Hillmich, Matthias Brandl, Robert Wille}
}

\maketitle

\begin{abstract}
    Trapped-ion quantum computers exhibit promising potential to provide platforms for \mbox{high-quality} qubits and reliable quantum computation. 
    The \emph{Quantum Charge Coupled Device (QCCD)} architecture is a leading example that offers a modular solution to enable the realization of scalable quantum computers, paving the way for practical quantum algorithms with large qubit numbers. 
    Within these devices, ions can be shuttled (moved) throughout the trap and through different dedicated zones, e.g.,~a memory zone for storage and a processing zone for the actual computation. However, due to decoherence of the ions’ quantum states, the qubits lose their quantum information over time. Thus, the required time steps of shuttling operations should be minimized. In this work, we propose a heuristic approach to determining an efficient shuttling schedule, which orchestrates the movement operations within the device. Given a quantum algorithm and a device architecture, the proposed approach produces shuttling schedules with a close-to-minimal amount of time steps for small-size QCCD architectures. 
    For large scale QCCD devices, empirical evaluations show promising results with respect to quality of the solution as well as performance. An implementation of the proposed approach is publicly available as part of the open-source \emph{Munich Quantum Toolkit}~(MQT,~\cite{willeMQTHandbookSummary2024}) at \url{https://github.com/cda-tum/mqt-ion-shuttler}.
\end{abstract}

\begin{IEEEkeywords}
quantum computing, design automation, ion shuttling, trapped ions
\end{IEEEkeywords}

\section{Introduction}
\label{sec:introduction}

\IEEEPARstart{Q}{uantum} computing promises to utilize fundamental quantum mechanical phenomena to tackle problems that are beyond the reach of classical computers. Famous example include Shor's algorithm to factorize integers~\cite{DBLP:conf/focs/Shor94}, Grover's search for unstructured data~\cite{DBLP:conf/stoc/Grover96}, as well as the simulation of quantum systems to advance the field of quantum chemistry~\cite{PhysRevX.8.011044}.

Over the past few decades, different physical platforms have been explored to realize quantum computers, such as superconducting quantum computers~\cite{Kjaergaard_2020}, neutral atom quantum computers~\cite{Henriet2020quantumcomputing, Bluvstein_2023}, or optical quantum computers~\cite{Slussarenko_2019}. Among these, trapped-ion quantum computers have emerged as a leading candidate due to their high-fidelity qubits and long coherence times. In a trapped-ion system, ions are confined and manipulated using electromagnetic fields, allowing precise control over qubit states and interactions. 

One of the main strengths of \mbox{trapped-ion} quantum computers is their ability to physically move ions in space.
The \emph{Quantum Charge Coupled Device} (QCCD) architecture proposes to exploit this to enhance the scalability and enable the construction of large-scale devices.
The ability to shuttle the ions within the device allows for \mbox{all-to-all} connectivity of the system's qubits, since all ions can be addressed by changing the routing of the qubits, i.e., moving the ions in such a way, that the respective ions are coupled.
Furthermore, individual trap regions can be optimized towards specific tasks, \mbox{e.g., for} processing or storing, between which ions are efficiently shuttled.

Despite these promises, several challenges remain in the development of practical and scalable quantum computers. One of the most pressing issues is the decoherence of qubits, i.e., the fact that the fragile quantum states lose information over time due to interactions with their environment. To address these challenges, interdisciplinary efforts are necessary, combining insights from quantum physics and computer science. From the perspective of computer science, the resulting expertise of decades of classical computing should be used to provide tools that compile, evaluate, and help in the development of new devices.
Without proper support, there is the possibility that powerful trapped-ion quantum computers will be available but there will be no means to use that power.
Indeed, this holds true for all quantum computing technologies.

In trapped-ion systems, especially in large architectures, a major part of the execution time may be consumed by shuttling operations.
Thus, a high-level quantum circuit has to be compiled in a way that not only permits the execution of the circuit's quantum gates, but also orchestrates the efficient movement of all ions in the system to ensure fast execution times.
This makes determining optimized shuttling schedules paramount for useful computations in trapped-ion quantum computers.
First solutions addressing this problem have been proposed, e.g.,~in~\cite{Schmale_2022, durandau2022automated, 9138945, Kreppel_2023, schoenberger2023using}. 
However, the considered architectures are comparatively simple and do not cover a wide range of possible QCCD architectures.

In this work\protect\footnotemark\footnotetext{Preliminary versions of this work have been published in \cite{schoenbergerCyclebasedShuttlingTrappedIon2024}.}, we propose a heuristic approach to determining efficient shuttling schedules for grid-type QCCD architectures.
To this end, the compilation process from a \mbox{high-level} quantum circuit to a precise shuttling schedule is divided into two parts:
first, we discuss the compilation of the quantum circuit to the specifics of the device. This process involves removing all SWAP gates, translating the high-level gate instructions to the native gate set of the device, optimizing the resulting circuit and mapping the logical qubits in the quantum circuit to the individual ions.
Second, we introduce a graph-based abstraction of the underlying physical hardware that represents linear ion traps and junctions.
Based on this graph, we generate efficient shuttling schedules without conflicts by exploiting cycles in the graph representation.
This enables movement on shortest paths without expensive backtracking to move potentially blocking ion chains out of the way.
To improve the efficiency of the resulting shuttling schedules, this step is intertwined with an additional compilation step that selects the next gate at runtime depending on the state of the system.

Empirical evaluations confirm the efficacy of the proposed approach with a close-to-minimal amount of time steps for small architectures and promising results for larger ones.
This includes both the resulting schedule (i.e.,~the number of time steps required to execute the quantum algorithm) as well as the classical generation of the schedule in the first place.
An implementation of the proposed approach is publicly available as part of the open-source \emph{Munich Quantum Toolkit}~(MQT) at \url{https://github.com/cda-tum/mqt-ion-shuttler} under the MIT~license.  

The remainder of this paper is structured as follows:
\mbox{\autoref{sec:background}} provides the background on trapped-ion quantum computers and QCCD architectures.
\autoref{sec:Shuttling for Scalable Trapped-Ion Quantum Computers} motivates the problem and outlines the general idea of the proposed solution.
\autoref{sec:gate-selection} details the additional compilation step that selects the next best gate.
\autoref{sec:path-generation} describes the proposed scheduling approach, with the corresponding implementation provided in \autoref{sec:implementation}.
\autoref{sec:evaluation} summarizes the obtained results. 
Finally, \autoref{sec:conclusion} concludes the paper.

\section{Background}\label{sec:background}
This section provides a brief overview of trapped-ion quantum computing and the challenges that have to be addressed to realize scalable devices. A potential solution, the \mbox{\emph{Quantum~Charge Coupled Device}} (QCCD) architecture, is explained and the costs associated with shuttling ions through a device are discussed.
For a more physics oriented description of the technology, the interested reader is referred to the provided references.

\subsection{Trapped-Ion Quantum Computing}
\label{sec:ion-trap-qc}

\begin{figure}[!t]
    \centering
    \subfloat[Paul trap]{\includegraphics[trim=0 -1em 0 0, clip, width=0.6\linewidth]{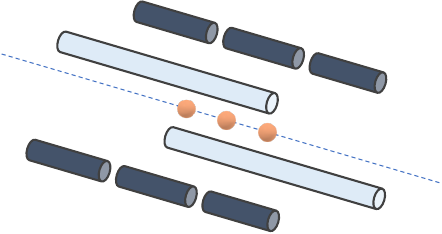}%
    \label{fig:single-linear-trap-3d}}
    \hfil
    \subfloat[Surface trap]{%
        \begin{minipage}{0.8\linewidth}
            \centering
            \includegraphics[trim=-8em 4.5em 12em 0, clip, width=.78\linewidth]{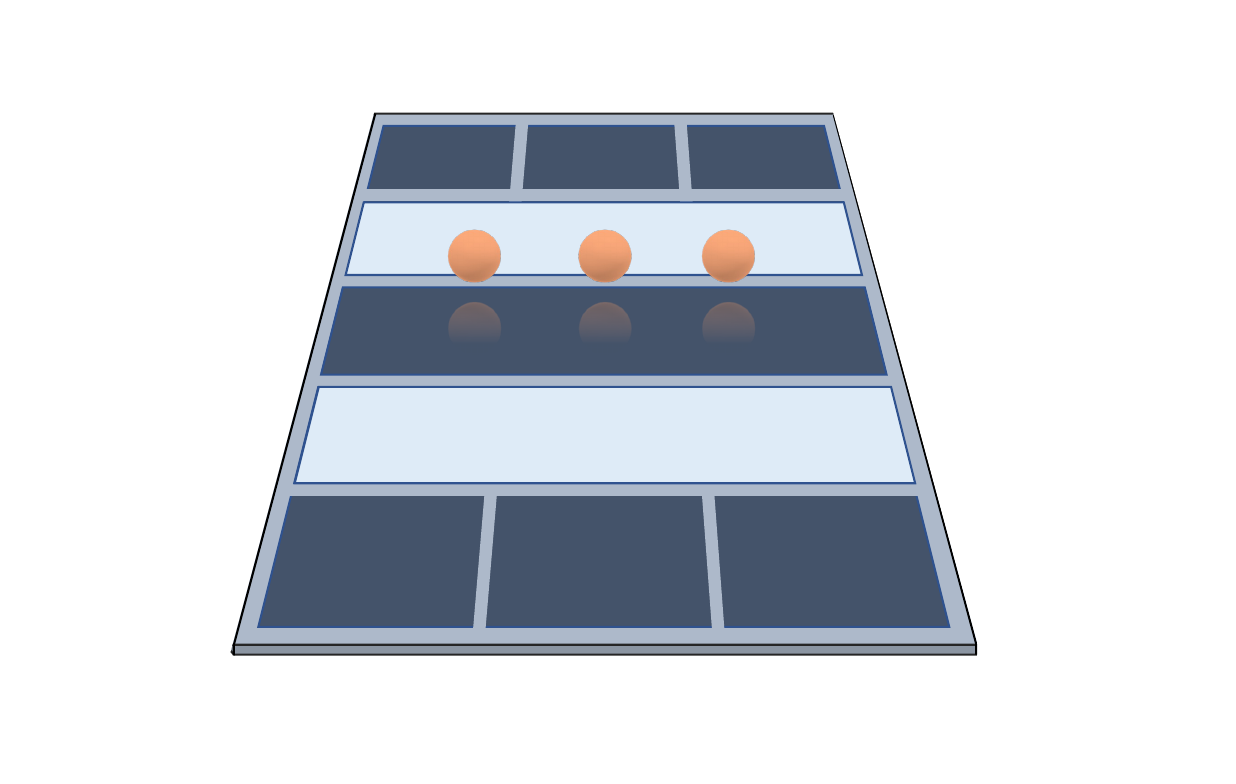}
            \includegraphics[trim=-2em -15em -1em 0, clip, width=0.2\linewidth]{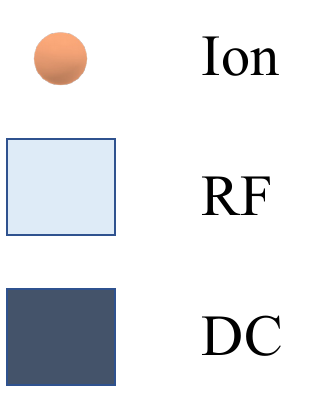}
        \end{minipage}%
        \label{fig:single-linear-trap-2d}}
    \vspace{.2em}
    \caption{An illustration of two possible linear trap realizations. A correct combination of \mbox{radio-frequency} (RF) and \mbox{quasi-static} (DC) electric fields produced by control electronics (\mbox{light-blue} and \mbox{dark-blue}) creates a potential that confines the ions (orange) at their current position.}
    \label{fig:single-linear-trap}
\end{figure}

Trapped-ion quantum computers~\cite{PhysRevLett.113.220501,PhysRevLett.74.4091,Debnath2016} utilize ions as qubits, where the quantum state of each ion is manipulated using electromagnetic interactions, either in the optical or microwave domain~\cite{PhysRevLett.74.4091, Harty_2016, Debnath2016}. To this end, ions are isolated and held in a controlled environment by a combination of \mbox{radio-frequency} and \mbox{quasi-static} electric fields. These fields generate an electric potential that confines the ions at a specific location.
Within the confines of a trap, multiple ions can be arranged in a \mbox{chain-like} configuration.

\begin{example}
    A popular type of an ion trap is the so called Paul trap.
    \autoref{fig:single-linear-trap} sketches a realization of a Paul trap that holds a single ion chain. Ions are held in an electric field generated by \mbox{radio-frequency} (RF, \mbox{light-blue}) and \mbox{quasi-static} (DC, \mbox{dark-blue}) control elements.
    The trap can also be fabricated as a \mbox{two-dimensional} surface trap as shown in \autoref{fig:single-linear-trap-2d}, which we refer to as a \emph{single linear trap site}.
    In both realizations, ions are illustrated by orange spheres representing three individual ions. This kind of an ion chain has been coined an \emph{ion register}, because the chains may be used similar to registers in classical computers. 
    \mbox{Gate-based} quantum computations can be performed on each ion individually, which means that each ion represents one qubit.

\end{example}

However, while a single trap suffices for smaller quantum computers, the gate speed~$R_{\textrm{gate}}$ decreases approximately by $R_{\textrm{gate}} \sim \frac{1}{\sqrt{N}}$ with an increasing number of ions $N$. Longer gate times give rise to different types of background errors, making it challenging to scale to more practical quantum algorithms that require more qubits.
As of now, \mbox{trapped-ion} quantum computers have been realised using up to tens of qubits~\cite{Brown2021, moses2023race}.

\subsection{Quantum Charge Coupled Device Architecture}
\label{sec:qccd}

An intuitive way of the limited scalability is to build systems with multiple ion chains. 
In fact, single linear trap sites can be connected to one large trap, which may hold one chain at each site. By exploiting the fact that ions can be physically moved in a trap, such modular architectures allow \mbox{all-to-all} connectivity of the system's ions. The leading candidate for a modular trap design is called the \emph{Quantum Charge Coupled Device}~(QCCD) architecture~\cite{Kielpinski2002}.  
The main idea of the QCCD architecture revolves around designating specific regions of the trap for specific functionalities. For instance, all quantum operations are performed in a dedicated \emph{processing zone} that is specifically constructed for efficient qubit operations.
The acquired quantum information may then be stored in a \emph{memory zone}, which is shielded from potential disturbances and sources of decoherence. Further optimized areas may include regions for qubit readout (\emph{measurement zone}) or ion initialization (\emph{loading~zone}).
Linear QCCD architectures have already been realized, e.g.,~in~\cite{Debnath2016,Pino2021}. A linear system is built from multiple sites, each able to confine one chain, connected in a straight line.

\begin{figure}[t]
    \centering
    \subfloat[Linear QCCD architecture]{\includegraphics[trim=0 -3em 0 0, clip, width=0.8\linewidth]{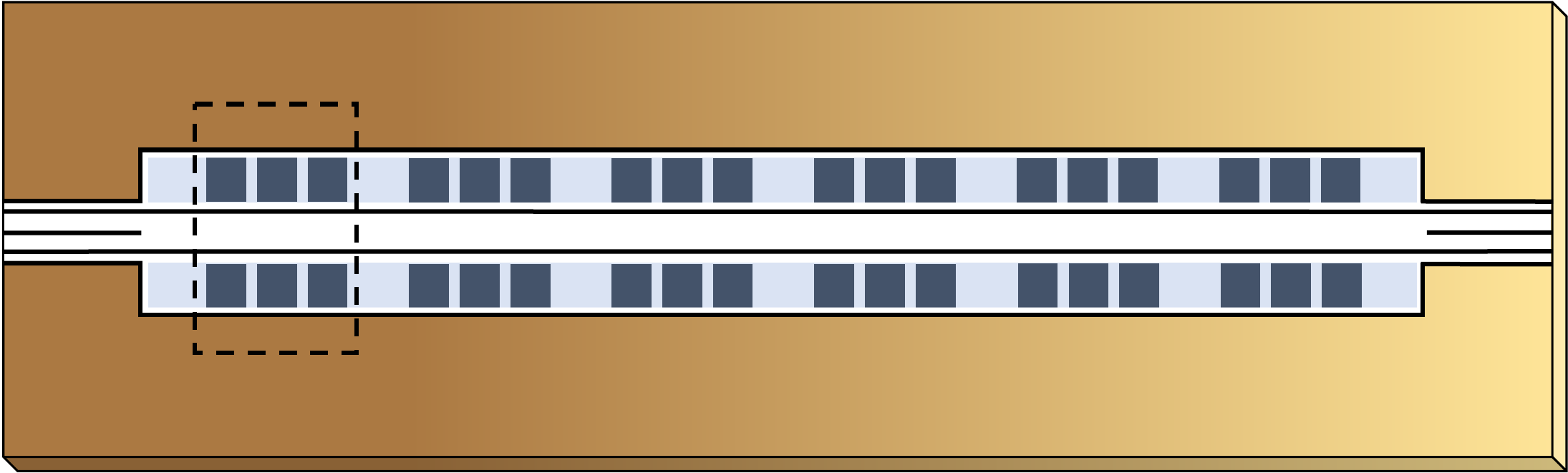}%
    \label{fig:linear_trap}}
    \hfil
    \vspace{1em}
    \subfloat[2D QCCD architecture]{\includegraphics[trim=0 -4em 0 0, clip, width=.92\linewidth]{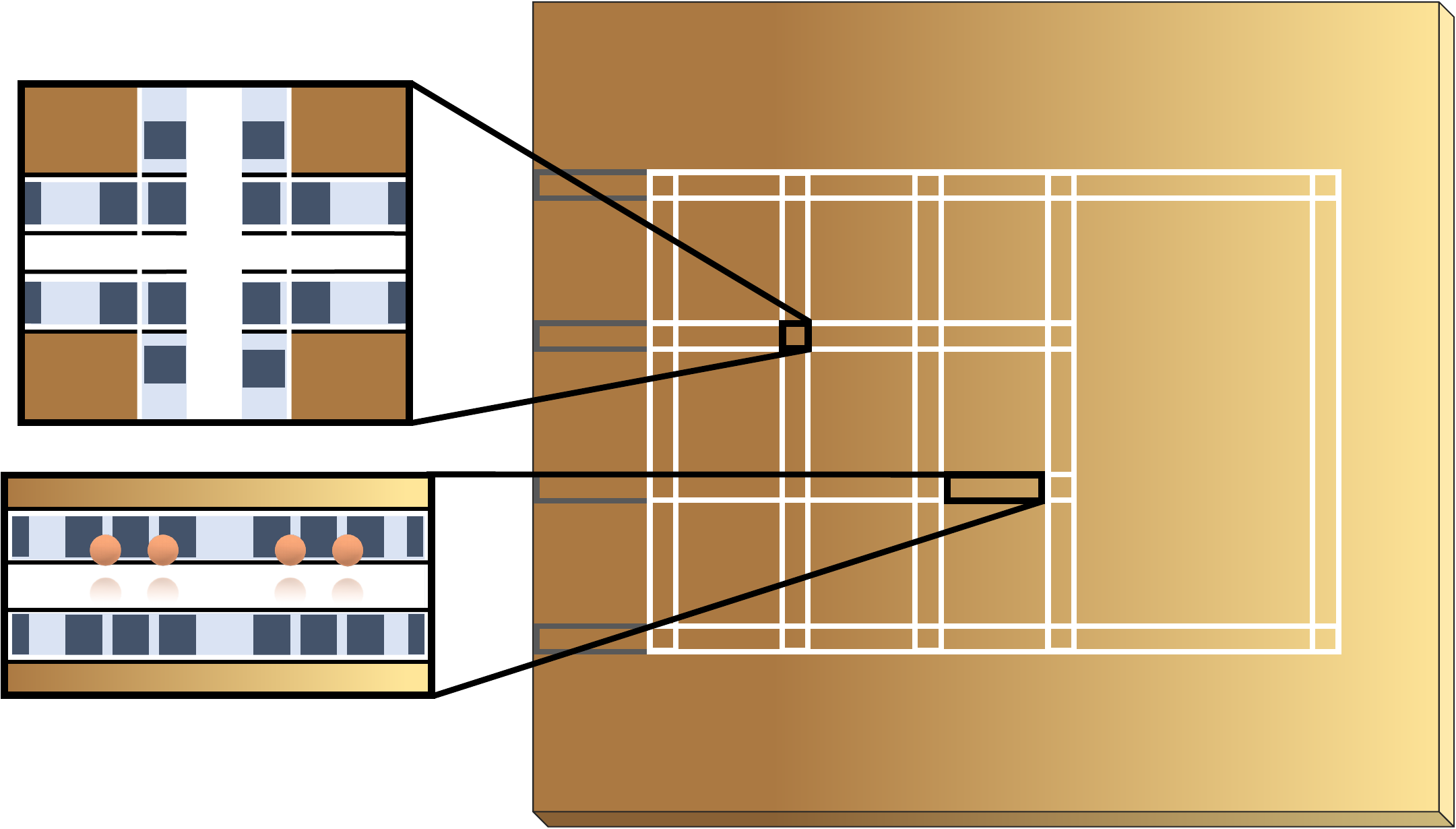}%
    \label{fig:qccd-example}}
    \vspace{.2em}
    \caption{An illustration of two QCCD architectures. In \protect\subref{fig:linear_trap} a device with a linear trap design is sketched. A 2D QCCD architecture is displayed in \protect\subref{fig:qccd-example}, wherein an \enquote{X}-junction and a linear region are indicated.}
    \label{fig:qccd}
\end{figure}

\begin{example}\label{ex:linear-trap}
    \autoref{fig:linear_trap} illustrates the concept of a linear QCCD device. 
    Each site is marked by three control electrodes (\mbox{dark-blue}), which trap the ions and may perform shuttling operations to move an ion register to a neighbouring site.
\end{example}

In linear traps, ion chains may block the way of each other, which would require slow interactions like chain reordering and reconfiguration to resolve this issue. To address this problem, junctions are implemented into the systems, connecting linear regions to form two-dimensional (2D) architectures. The extension to a second dimension allows ions to avoid the path of other ions without swapping or reconfiguration. Linear regions can be connected in several ways. One may connect large linear regions with only a small amount of junctions, or increase the number of junctions and connect several smaller linear regions.
Furthermore, the combination of different types of junctions (e.g.,~termed \enquote{T}-, \enquote{Y}- or \mbox{\enquote{X}-junctions}, where the capital letter is referring to their shape) can also completely change the layout and connectivity of the system.
A first type of a two-dimensional QCCD device has recently been realized in~\cite{moses2023race}, where a linear trap was connected to form a loop.

\begin{example}
    \autoref{fig:qccd-example} shows one possible layout of a 2D QCCD device. 
    All inner linear regions are connected via \mbox{\enquote{X}-junctions}, which produces a system in a grid structure. Each linear region can hold up to two chains. 
    In this architecture, the square grid on the left-hand side is dedicated as a memory zone which is connected to a linear processing zone on the right-hand side.
\end{example}

\subsection{Shuttling Costs}

Conducting the shuttling operations of ion chains in a QCCD architecture comes with certain costs.
Depending on the architecture, different amounts of energy have to be invested into the system to trap and move the ions. 
The underlying physics allow comparatively fast shuttling through linear traps, however, shuttling through junctions is more difficult and dominates the overall time needed for moving ion chains~\cite{Wright_2013}.
Over time (and through the movement), the ions collect energy through acquisition of phonons~\cite{arxiv.1702.02583}, resulting in a higher likelihood of errors and decoherence.
Therefore, ions must be cooled to or close to the ground state---often by a combination of Doppler and side-band cooling (see~\cite{Bruzewicz_2019} for more information) to preserve their quantum information.
It still remains challenging to integrate the required optical control elements, especially for large systems. 
To limit the amount of control elements needed and enable the development of large-scale devices, it is therefore crucial to find shuttling schemes that minimize the time ions have to be moved through the system.
    
\begin{figure*}[!h]
    \centering
    \subfloat[Quantum circuit\protect\footnotemark]{\includegraphics[trim=0 -4em 0 0, clip, width=.3\linewidth]{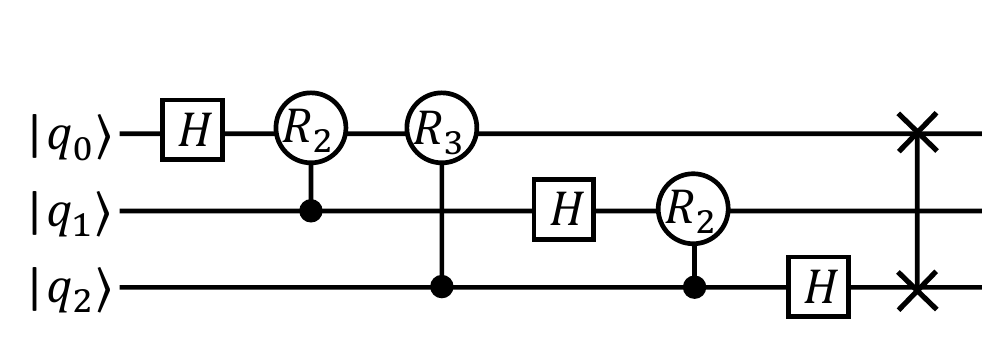}%
    \label{fig:flow_circuit}}
    \hfil
    \subfloat[QCCD device]{\includegraphics[trim=-2cm -.5em -2cm 0, clip, width=.3\linewidth]{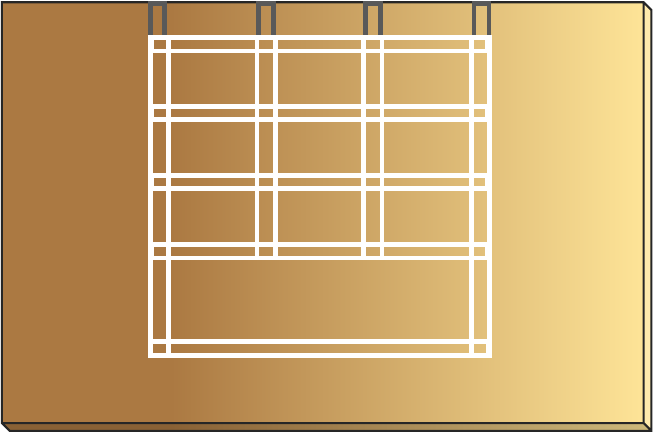}%
    \label{fig:flow_qccd}}
    \hfil
    \subfloat[Starting configuration]{\includegraphics[trim=0 -.5em 0 -.5cm, clip, width=.3\linewidth]{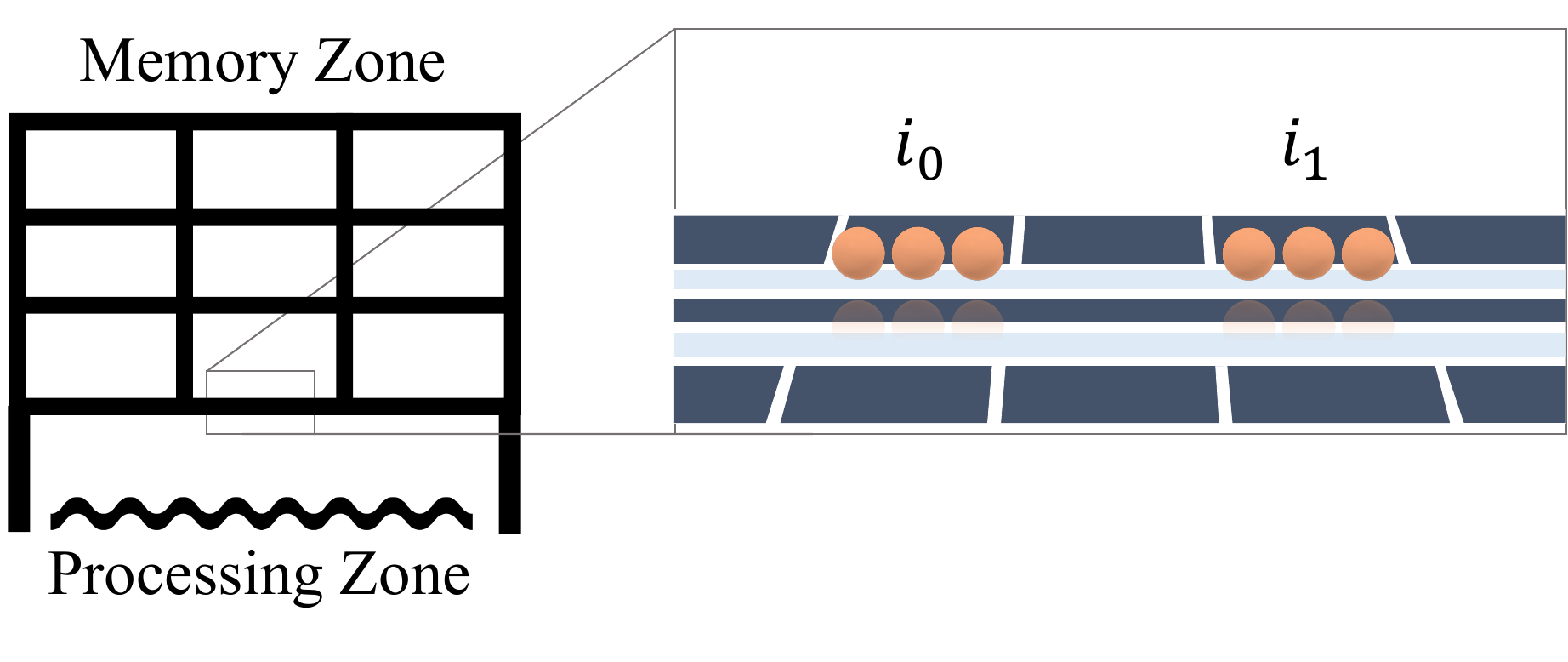}%
    \label{fig:flow_initial}}
    \hfil
    \vspace{.5em}
    \subfloat[Compilation flow]{\includegraphics[trim=0 -.5em 0 0, clip, width=\linewidth]{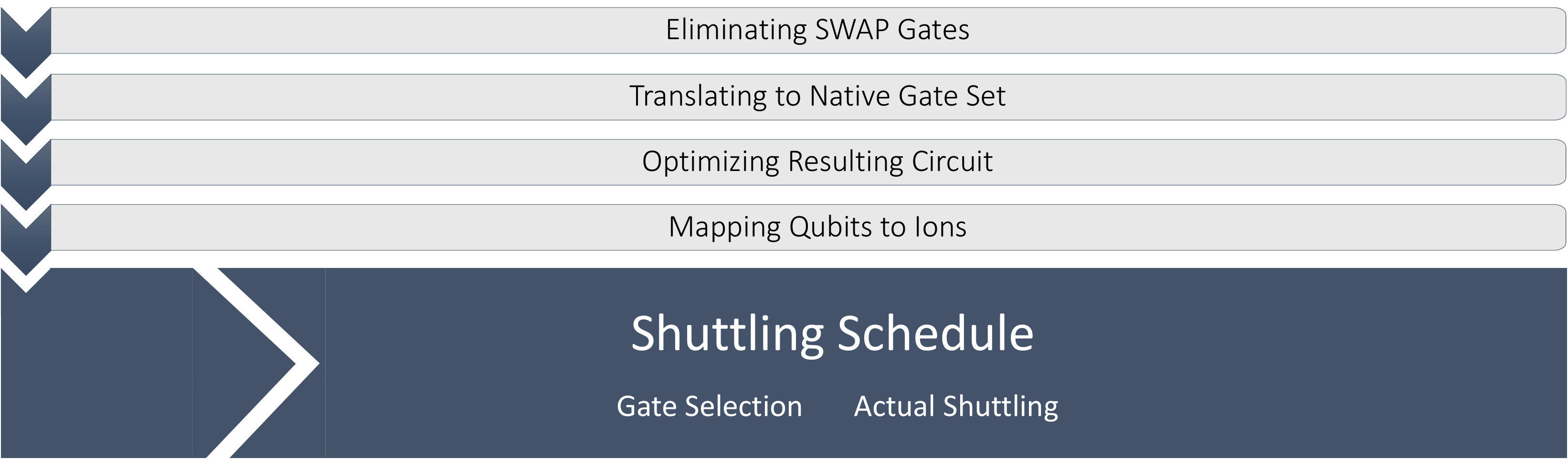}%
    \label{fig:flow_below}}

    \caption{Input and intermediate steps for the generation of a shuttling schedule. The top layer shows the necessary input for the proposed approach. Below, all consecutive steps of the compilation that generate the final shuttling schedule are displayed.}
    \label{fig:flow_chart}
\end{figure*}

\section{Shuttling for \\Scalable Trapped-Ion Quantum Computers}
\label{sec:Shuttling for Scalable Trapped-Ion Quantum Computers}

This section reviews the shuttling problem specific to trapped-ion quantum computers, that is abstracted and solved in the following sections.
To this end, we first provide a comprehensive overview of the compilation flow that prepares a \mbox{high-level} quantum circuit to be executed on actual hardware and, with that, provide the necessary context.
In the second part, we then describe the steps involved in the shuttling process and the corresponding problem instance. Based on that, the remainder of this work proposes an automatic solution for this problem.

\subsection{Compilation Flow and Context}

To provide a consistent description of the problem, we start with a review of the compilation process that leads up to the shuttling problem.
At the start of every computation, the compiler is given a quantum circuit, usually as a \mbox{high-level} description, i.e., consisting of quantum gates that the machine is not necessarily able to execute directly. 
Besides that, the compiler must also know the specific architecture of the considered QCCD device with its corresponding starting configuration.

\begin{example}
    The compilation flow is illustrated using \autoref{fig:flow_chart}. Here, the compiler is given the quantum circuit of the \mbox{3-qubit} quantum Fourier transform~\cite{DBLP:books/daglib/0046438} illustrated in \autoref{fig:flow_circuit} as input. Further input to the compiler consist of information about the employed QCCD device, namely the type of architecture and number of ion chains, as well the position of the ion chains at the start of the execution, which are depicted in \autoref{fig:flow_qccd} and \autoref{fig:flow_initial}.
\end{example}

As mentioned before, the QCCD device may not be able to natively run the \mbox{high-level} description of the quantum circuit. To resolve this, the circuit has to be compiled to satisfy the restrictions of the architecture. For the specifics of a \mbox{shuttling-based} trapped-ion hardware, we break the compilation down into the following four steps:

\begin{itemize}
\item \emph{Eliminating SWAP Gates}\\
    Corresponding to its name, a SWAP gate swaps the considered qubits of the gate within the circuit. A shuttling based quantum computer is able to realize swapping of qubits by simply relabeling the qubits since the shuttling allows connectivity between all qubits. As a consequence, the compiler first eliminates all SWAP gates from the original circuit and relabels the qubits accordingly.
\item \emph{Translating to Native Gate Set}\\
    A high level quantum circuit is usually constructed by abstract quantum gates, that are not necessarily executable on a quantum machine. Each realization of a quantum computer only supports a very specific set of quantum gates, called  the \emph{native gate set}. Accordingly, the compiler has to translate the high level description of a quantum circuit into the native gate set of the considered quantum device.
\item \emph{Optimizing Resulting Circuit}\\
    Translating a quantum circuit to the native gate set usually introduces multiple gates for each gate of the original circuit. The task of a compiler is then to reduce this overhead and, thus, minimize the amount of shuttling that has to be done later to execute the circuit on the machine. To achieve this, the compiler removes redundant gates or gate sequences.
    One can also make use of the fact that quantum gates may commute and change the execution order of gates in a way that is favorable for later stages of the shuttling compilation or introduce more redundancies that can be removed.
\item \emph{Mapping Qubits to Ions}\\
    The initial configuration of ion chains provides the location of each ion chain within the device and the amount of ions that each chain holds. The compiler maps the qubits of the quantum circuit (logical qubits) to the individual ions that act as physical qubits.%\\
\end{itemize}
\vspace{.3em}

\footnotetext{$R_n$ gates are defined by $R_n=\begin{psmallmatrix} 1 & 0 \\ 0 & e^{2\pi i/{2^n}} \end{psmallmatrix}$.}

The final step above is essential for the shuttling process considered in this work. In the remainder of this work, we are using the following terminology.

\vspace{.5em}
\begin{definition}
        We denote the set of ions as \mbox{$I = \{i_0, \cdots, i_{m}\}$}. Multiple individual ions can be hold in one ion chain from the set \mbox{$C = \{c_0, \cdots, c_{l}\}$}, where each ion carries the information for one qubit. The chains also do not share ions, i.e., \mbox{$\forall_{i,j}:i \neq j\rightarrow c_i \cap c_j = \emptyset$}.
\end{definition}
\vspace{.5em}

Overall, the steps reviewed above are illustrated using the following example.

\vspace{.3em}
\begin{example}
    Consider again the example covered before using \autoref{fig:flow_chart}. To execute the quantum Fourier transform on three qubits, the circuit has to be prepared and mapped to the architecture of the provided quantum device. The compiler first removes the SWAP gates at the end of the circuit. Since the SWAP gates are not in between other gates, no relabeling of gates has to be done. As a next step, the compiler translates each gate of the original circuit into the set of gates that are native to the QCCD device, removes redundancies introduced by the mapping, and executes possible circuit optimizations.
    The first part of the resulting circuit is given in \autoref{fig:QFT_native_short}. 
    This part of the compilation is then finished by mapping the logical qubits of the circuit to the physical qubits of the device. In this case, we map the circuit in the most simple way, i.e., mapping the first qubit to the first ion $i_0$, the second qubit to the second ion $i_1$, and the third qubit to the third ion $i_2$.
    \label{ex:flow}
\end{example}
\vspace{.1em}
    After completing these first four steps, the quantum circuit is compiled to the native gate set of the device and mapped to the corresponding physical qubits. In principle, the actual execution of the quantum circuit and, with that, the shuttling, can start.  

\subsection{Shuttling}
    The role of a compiler for QCCD devices is paramount to ensure the effective shuttling of ions between different system zones.
    Fundamentally, quantum gates can only be operated in the processing zone. As a consequence, it is essential that qubits on which operations shall be employed are shuttled (moved) through different zones (e.g.,~from the memory zone where they are stored to the processing zone).

    With an executable circuit given, the order in which the ions have to be shuttled to the processing zone can in principle be determined.
    Since the possible movement inside the processing zone is immediately given by the architecture, we put the focus of this paper on the movement inside the memory zone and at the interface to the processing zone.
        
    To approach this in an efficient way, we separate the problem into two intertwined parts, one selecting the next gate, and one generating the paths for all ion chains to execute the respective gate, i.e., orchestrating the actual shuttling of the ions:

\vspace{.5em}
\begin{itemize}
    \item \emph{Gate Selection}\\
        The idea is, analogous to the previously discussed circuit optimizations, to exploit commutation relations of quantum gates. More precisely, because of the commutation of gates, multiple gates may be executable as the next gate, which allows the compiler to choose the most favorable one for the construction of a shuttling schedule. This is discussed in more detail in \autoref{sec:gate-selection}.
    \item \emph{Actual Shuttling}\\
        Once all previous steps are complete, the next gate and corresponding ion chains, which are needed in the processing zone, are determined.
        With that, a schedule of all movement operations can be generated.
        The compiler must orchestrate the shuttling paths of ion chains to and from the processing zone as well as the repositioning of other, potentially blocking, ion chains within the memory zone to enable efficient shuttling.
        This schedule should not only ensure the correct execution but also aim to minimize the total execution time of the quantum circuit.
\end{itemize}
\vspace{.5em}

Solving this shuttling problem is the main topic of this work and will be discussed in detail in the next two sections, i.e., \autoref{sec:gate-selection} will cover the \emph{Gate Selection} part, while \autoref{sec:path-generation} will cover the \emph{Actual Shuttling} part.

\section{Gate Selection}
\label{sec:gate-selection}

The core of the \emph{Gate Selection} part lies in a strategic selection of subsequent gates, particularly prioritizing those with ions nearest to the processing zone.
Since gates of a quantum circuit may commute, there exist multiple possible ways that execute the circuit without affecting the overall outcome of the quantum computation. In other words, the order of gates is not necessarily fixed at the start of the execution.
To exploit this, we first translate the quantum circuit to a so called \emph{Dependency Graph}. This allows us to isolate a \emph{front layer} of gates, i.e., gates that are ready to be executed next. With that, we are able to choose the gate from the front layer that is optimal for the shuttling process.

\subsection{Dependency Graph}
Every quantum circuit can be represented as a \emph{Directed Acyclic Graph} (DAG). Each gate in the circuit constitutes a node of the graph, with
directed edges between the nodes corresponding to gate dependencies, i.e., lack of commutation. The resulting graph is referred to as a \emph{Dependency Graph} of the respective quantum circuit. A directed edge from node $A$ to node $B$ means that gate $A$ does not commute with gate $B$, and thus, has to be executed beforehand. Therefore, the set of predecessors of a node include all gates that must be executed before the respective gate can be considered.

\subsection{Front Layer}
To exploit the fact that gates commute, we use the Dependency Graph to delineate a \mbox{\emph{front layer}} of gates. The front layer consists of all gates without preceding dependencies. As a result of the construction of the Dependency Graph, all gates in the front layer commute with each other, do not depend on the execution of other gates and, therefore, can be executed as the next gate.
This way, we can choose the most favorable one from the front layer to be the next gate in the execution order.
As such, gates acting on ions closest to the processing zone are executed first, allowing the immediate execution of certain gates and significantly reducing the necessity for extensive ion shuttling.

\begin{figure}
    \centering
    \subfloat[Circuit]{\includegraphics[trim=-1cm 0 0 0, clip, width=\linewidth]{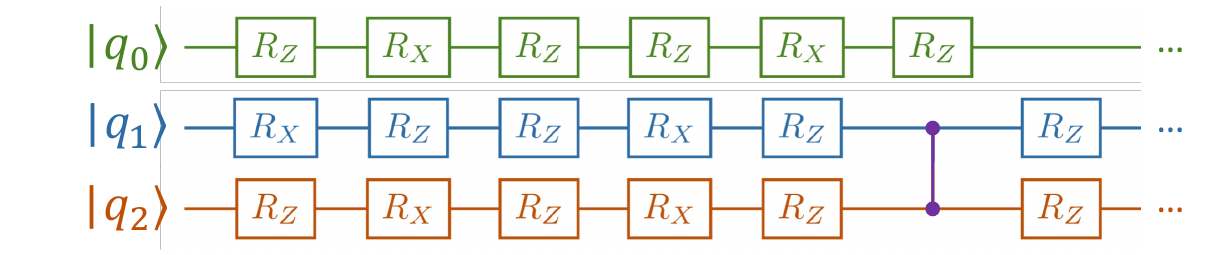}%
    \label{fig:QFT_native_short}}
    \hfil
    \vspace{1em}
    \subfloat[Dependency Graph]{\includegraphics[width=\linewidth]{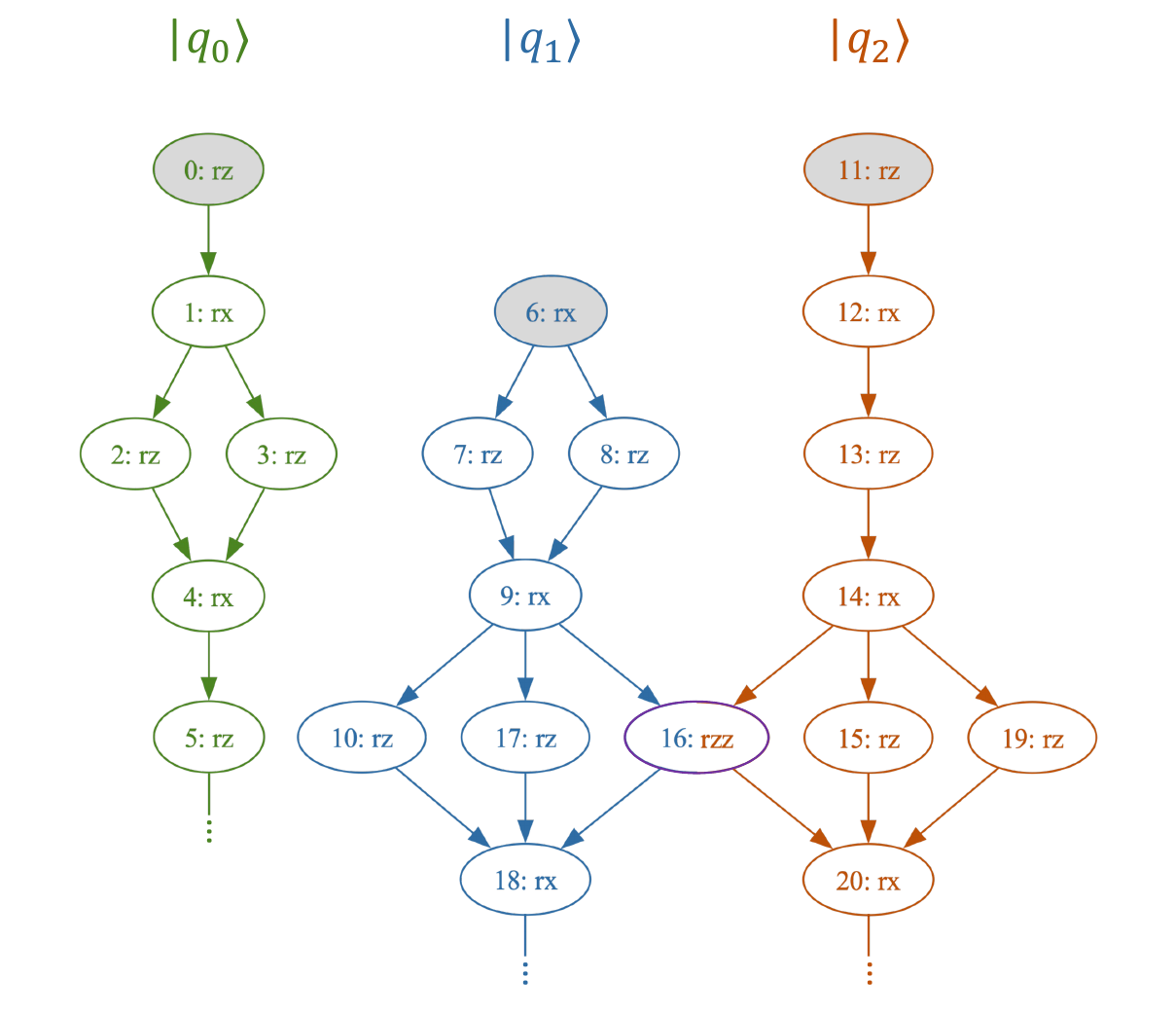}%
    \label{fig:dag_dep}}
    \caption{Circuit and Dependency Graph of the compiled \mbox{3-qubit} quantum Fourier transform. The front layer of the Dependency Graph is indicated (grey).}
    \label{fig:circuit_and_dag}
\end{figure}

\begin{example}\label{ex:dag}
Recap the state of the compilation flow at the end of \autoref{ex:flow}. The 3-qubit quantum Fourier transform given in \autoref{fig:flow_chart} is compiled to the respective QCCD device and its qubits are naively mapped to the ions.
The first part of the compiled circuit is displayed in \autoref{fig:QFT_native_short}.
Given this circuit, the corresponding Dependency Graph in \autoref{fig:dag_dep} can be constructed. The front layer of gates is given by the nodes without preceding dependencies or in other words, the nodes to which no edges are directed towards. Consequently, the front layer consists of node $0$, $6$, and $11$.
\end{example}

\subsection{Best Gate}
Since all gates in the front layer are qualified to be the next gate in the execution order, we may pick the gate that is optimal for the shuttling scheme. The shuttling schedule consequently evolves around the chosen gate, as the ions of this gate are the ones that have to be moved to the processing zone next. Since we consider a shuttling based device, a favorable gate requires qubits that are close to or, at best, already present in the processing zone. In this approach, the gate with the minimal combined distance to the processing zone of its qubits is considered the best gate.

\begin{example}\label{ex:qft_best_gate}
Consider again the circuit of the quantum Fourier transform on 3 qubits provided in \autoref{fig:flow_circuit}. The corresponding Dependency Graph is given in \autoref{fig:circuit_and_dag}. Let ion $i_1$ be closest to the processing zone. Since qubit $q_1$ has been mapped to $i_1$ before and node $6$ of the front layer is acting onto that qubit, this node and the respective $R_X$ gate on qubit $q_1$ is chosen as the next gate. This means that the shuttling schedule now produces shuttling operations to move the ion chain holding ion $i_1$ to the processing zone.
\end{example}

After the successful shuttling of the ions and execution of the respective gate, the node of this particular gate is removed from the Dependency Graph. With that, a new front layer is created and the most favorable next gate can be chosen depending on the new state of the ions in the system. This is repeated until the complete quantum circuit is executed. To efficiently manage the ion chains in such a way, that the selected ions arrive at the processing zone, we discuss an approach to generate paths and manage all ions of the system in the next section.

\section{Actual Shuttling}
\label{sec:path-generation}
With the completion of the compilation steps above, the compiler now has access to a compiled quantum circuit and, crucially, a Dependency Graph that determines which ions have to be in the processing zone next. With this information, a shuttling schedule has to be created that orchestrates the movement operations of all ions in the device, such that the given quantum circuit is executed.
An efficient shuttling schedule moves ions in a way that minimizes the overall time to execute the given circuit.
We propose to construct an efficient shuttling schedule by first abstracting the architecture of a QCCD device as an undirected graph and, afterwards, exploiting cycles in that graph to guarantee conflict-free movement. To provide a complete description for the discussed architecture, this section also covers the shuttling through a processing zone.

\subsection{Graph Description}

To find an appropriate shuttling schedule, we represent the architecture of a memory zone and the interface to the processing zone in a QCCD device as an undirected graph.
The edges of the graph represent the individual sites of linear traps (each site holding one ion chain).
The nodes represent either junctions (termed major nodes) or connections between sites in one linear trap (termed minor nodes).
On this graph, the physically continuous movement of ion chains is discretized into time steps.
At every time step, each chain is present at exactly one edge of the graph.
With this representation, the task at hand becomes a combinatorial optimization problem.

\begin{definition}\label{def:notation}
    Consider a graph $G = (V, E)$ that represents the architecture of the memory zone and the interface to the processing zone on the QCCD device.
    
    The set of nodes $V$ contains two different types: \emph{major nodes} representing junctions and \emph{minor nodes} that separate adjacent individual sites.
    
    The set $E = \{e_0, \dots, e_k\}$ denotes the edges of the graph, representing all possible positions of the ion chains.
    There are two special edges: the (i)~\emph{outbound edge} for ion chains exiting the memory zone for processing in the processing zone and the (ii)~\emph{inbound edge} for ion chains returning to the memory zone. 
    Edges can connect two major nodes (when there is only one site between junctions), a major and a minor node, as well as two minor nodes.
\end{definition}

Since, in this particular architecture, individual ions within an ion chain are not directly reconfigured or split from their chain, ion chains represent the individual objects that are managed by the shuttling schedule. While the previous discussions can be adapted to the scheduling of individual ions, from this point on, this work focuses on the movement of ion chains instead of individual ions.

\begin{figure}
    \centering
    \subfloat[QCCD device]{\includegraphics[trim=-1cm 0 24cm 0, clip, width=.48\linewidth]{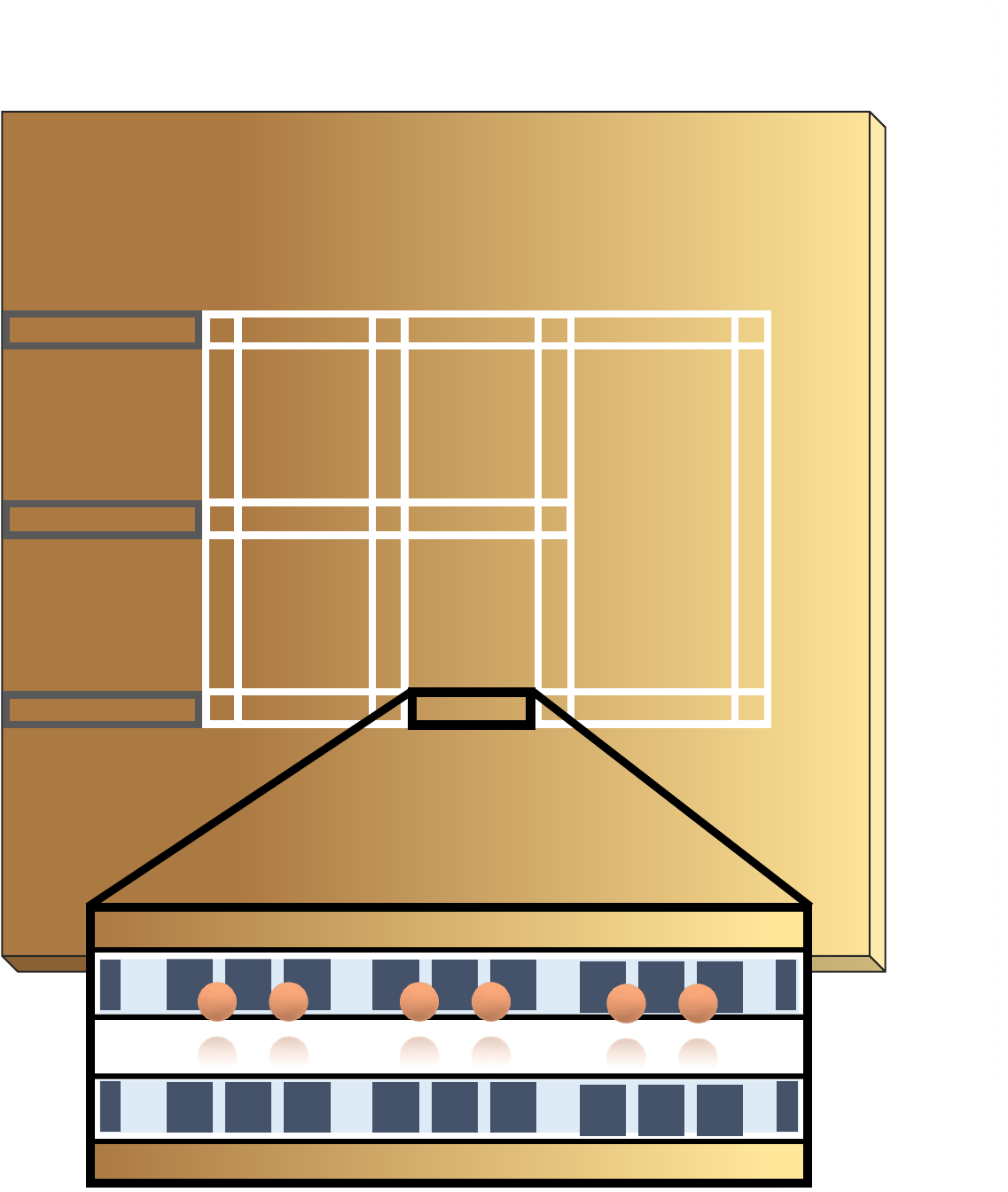}%
    \label{fig:QCCD_device}}
    \hfil
    \subfloat[Graph]{\includegraphics[trim=20cm -2cm 0cm 0, clip, width=.48\linewidth]{Figures/QCCD_Graph.pdf}%
    \label{fig:graph}}
    \caption{Illustration of a 2D QCCD architecture and its corresponding graph abstraction}
    \label{fig:QCCD_and_Graph}
\end{figure}

\vspace{.3em}
\begin{example}
    Consider the QCCD architecture in \autoref{fig:QCCD_device}.
    All linear regions between two junctions can hold up to three individual ion chains. 
    A corresponding graph is illustrated next to the device where the major nodes (green) form a grid of size $3\times3$.
    Minor nodes (black) mark the three individual sites between two junctions. One inbound edge leads to the edge of the processing zone, which is connected back to the memory zone by one outbound edge. The inbound and outbound edge connecting the processing zone to the memory zone are marked in the graph of \autoref{fig:graph}.
\end{example}
\vspace{.3em}

On the graph representation, the state of the memory zone is fully described at any time step.
At time step zero, each ion chain is located at the edges according to the starting configuration.
Given the information of which chains are needed in the processing zone, the next step in the compilation process is to find a schedule of shuttling operations, which efficiently moves the respective chains.
Moving an ion through a junction takes considerably more time time than moving withing a linear region.
Thus, one time step passes whenever an ion is moved through a junction.
While one ion chain is shuttled through a junction, all other junctions may shuttle other ions at the same time.
This means all junctions can be used in parallel within one time step.
An example illustrates the concept.

\begin{example}
    An ion chain configuration of the memory zone is illustrated in \autoref{fig:next-move-0}. In a naive approach, all chains (black boxes) travel on their shortest path, given they are needed in the processing zone as dictated by the Dependency Graph. For each chain, the shortest path to the processing zone is indicated in \autoref{fig:next-move-0}.
\end{example}

\begin{figure}
    \centering
    \subfloat[Individual]{\raisebox{0.2cm}{\includegraphics[width=.5\linewidth]{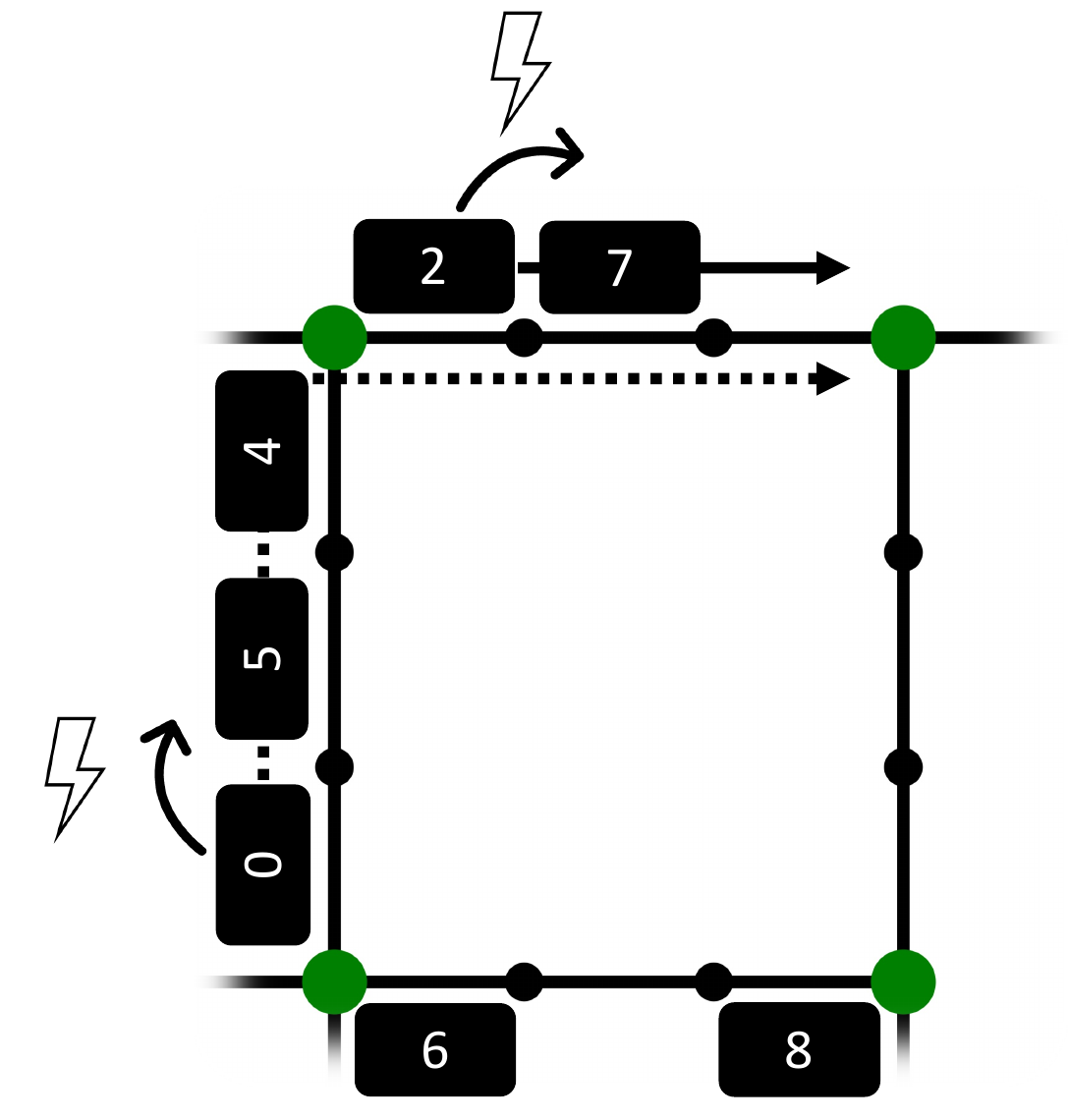}}%
    \label{fig:actual_initial}}
    \hfil
    \subfloat[Cycle]{\includegraphics[width=.45\linewidth]{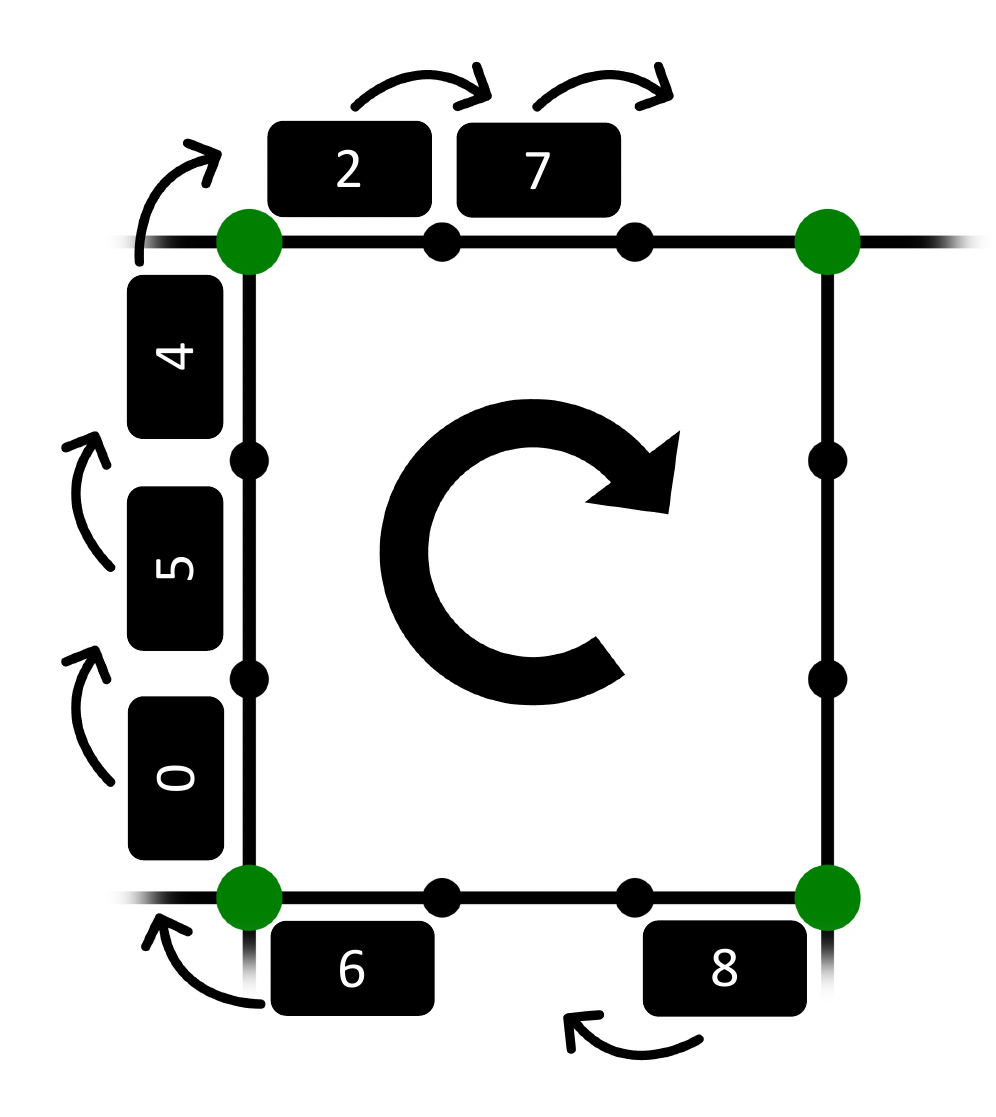}%
    \label{fig:actual_circle}}
    \caption{Comparison of individual consideration of ion chains with conflicts and conflict-free movement with cycles}
    \label{fig:circle}
\end{figure}

\subsection{Cycle-based Shuttling}

The defining problem of shuttling arises if an ion chain meets other chains on its path. 
In a trap filled with multiple ion chains, the shortest path to the processing zone may be blocked, since the chains can not directly swap places in a memory zone.
These conflicts have to be resolved by moving the blocking chains away from the paths or change the path of the moving chains. 
For an increasing number of ion chains in the system this problem becomes increasingly more difficult to solve exactly.

\vspace{.3em}
\begin{example}
    Consider \autoref{fig:actual_initial}, which displays one rectangle within a grid graph, similar to the graph given in \autoref{fig:graph}. 
    When trying to move the ion chains $c_0$ and $c_2$, additional ion chains block the shortest paths to the processing zone. 
    The resulting conflicts are indicated, as chain $c_5$ ($c_7$) is blocking the way of chain~$c_0$~($c_2$).
\end{example}
\vspace{.3em}

To tackle this issue, the topology of the considered modular QCCD architectures offers an intuitive solution: exploiting \emph{cycles}.
Cycles avoid conflicts and still move chains on their shortest path within the memory zone.
On the graph representation, we refer to cycles as connected edges that form closed loops.
If we form cycles along the shortest paths of the chains and move every chain one step on that cycle, we are able to shuttle individual chains along their optimal path while, at the same time, moving blocking chains away from that path.
Additionally, because all junctions can be shuttled through in parallel, one turn of all edges on a cycle only takes one time step.

\vspace{.3em}
\begin{example}
    Revisit the conflicts depicted in \autoref{fig:actual_initial}. Moving all chains on a cycle around the shown part of the graph, both $c_0$ and $c_2$ are able to move on their respective shortest paths. To do that, we move all chains on that cycle one edge further in the same direction. This is illustrated in \autoref{fig:actual_circle}.
\end{example}
\vspace{.1em}

To better illustrate the concept of cycle-based shuttling, we discuss the approach tailored to grid-type architectures from this point on.
Grid-type architectures are connected only by \enquote{X}-junctions, i.e.,~the angles of all junctions are \SI{90}{\degree}. 
This means, that each face of the graph is a rectangle (i.e.,~bounded by edges between four major nodes).
This shape offers a straightforward construction of cycles, since every rectangle of the grid forms a closed loop.
Depending on the direction of the desired movement, two different cycles are being constructed.

\begin{figure}
    \centering
    \subfloat[In memory zone]{\includegraphics[width=.42\linewidth]{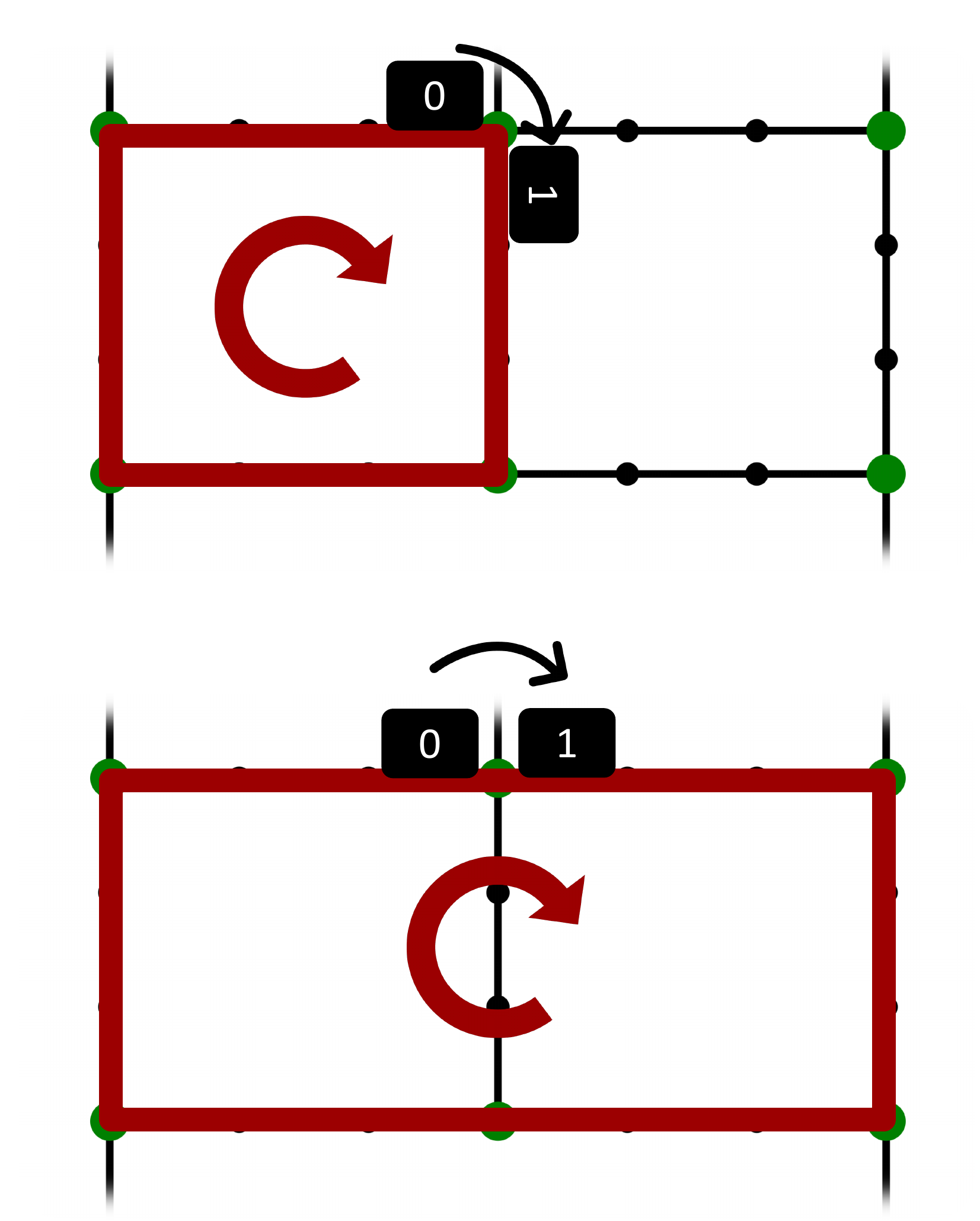}%
    \label{fig:circle_construct_left}}
    \hfil
    \subfloat[Through processing zone]{\raisebox{0cm}{\includegraphics[trim={1cm, 0, 0, 0}, width=.58\linewidth]{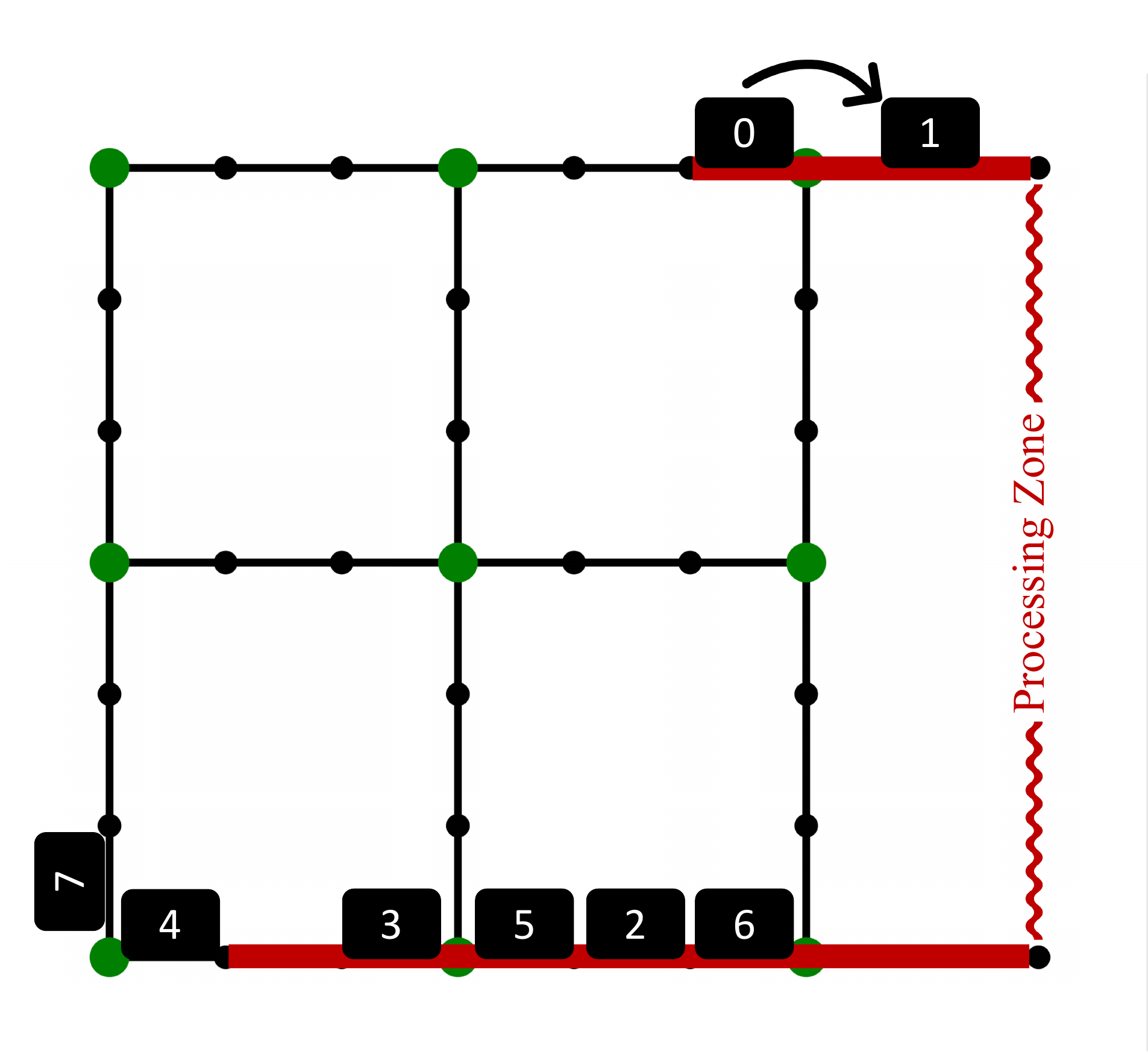}}%
    \label{fig:circle_construct_right}}
    \caption{Cycle construction}
    \label{fig:circle_construction}
\end{figure}

\vspace{.3em}
\begin{example}
    Since we consider a grid built out of \mbox{\enquote{X}-junctions}, a movement through a junction can either be horizontal or vertical. The smallest possible cycle for a vertical move requires exactly one rectangle of the grid, while for the horizontal move the cycle has to be expanded to two rectangles. These moves are exemplified in \autoref{fig:circle_construct_left}, in which ion chain $c_0$ is blocked by ion chain $c_1$ both times, once vertically~(top) and once horizontally~(bottom).
\end{example}
\vspace{.3em}

All move operations within the memory zone are covered by these two cycles, since the memory zone is a symmetric grid. To cover the shuttling operations for the whole architecture, the shuttling through the processing zone is discussed next.

\subsection{Shuttling through the Processing Zone}
In this work, we consider the processing zone to be a \mbox{one-way} pass that is separated from the \mbox{grid-type} memory zone. Constructing cycles through this pass is undesirable, because this may unintentionally move neighbor chains into or out of the processing zone. 
This issue is resolved by creating a path through the processing zone towards an unoccupied edge in the memory zone. This way, all chains on that path may move one edge further, which enables specific movement into and out of the processing zone. 
The start point for searching a free edge depends on where the chain is needed after being processed. If the chain is needed in the processing zone again at a later point in time, a breadth-first search starts at the inbound edge, otherwise we employ the same search starting from the edge with the largest distance to the inbound edge.

\vspace{.3em}
\begin{example}
    Consider the state of the memory zone in \autoref{fig:circle_construct_right}. Chain $c_0$ is on its way to use the inbound edge to move to the processing zone. The inbound edge is already occupied by $c_1$. Given $c_0$ is not needed another time, a breadth-first search for a free edge is performed starting from the bottom left corner node, opposite to the inbound edge. The resulting path from edge of $c_0$ to the unoccupied edge is indicated in \autoref{fig:circle_construction}.
\end{example}
\vspace{.3em}

Besides the movement through the processing zone, a shuttling schedule must also account for the time ions remain in the processing zone.
The speed of quantum gates in trapped-ion quantum computers is a focal point of ongoing research and development efforts. While current implementations demonstrate notable progress~\cite{Saner_2023,weber2024robust,nünnerich2024fast}, the execution time of quantum gates still remains a major part of the overall execution time.
Notably, two-qubit gates typically exhibit slower operation speeds compared to single-qubit gates.
As a consequence, a compiler for trapped-ion quantum computers must be able to schedule different time slots, in which ions remain in the processing zone. This may also temporarily block the path from the memory zone to the processing zone, in order to fully execute each gate.

As already discussed, ions have to be shuttled to the processing zone in a specific order to perform the calculations of a quantum circuit. 
To explain   how to preserve the order of the ions and decide which cycles to construct and execute, details of the corresponding implementation are discussed in the following section.

\section{Implementation}
\label{sec:implementation}

\begin{figure*}[t]
    \centering
    \subfloat[Start of time step]{\includegraphics[trim=0cm -1em 70cm 0, clip, width=.32\linewidth]{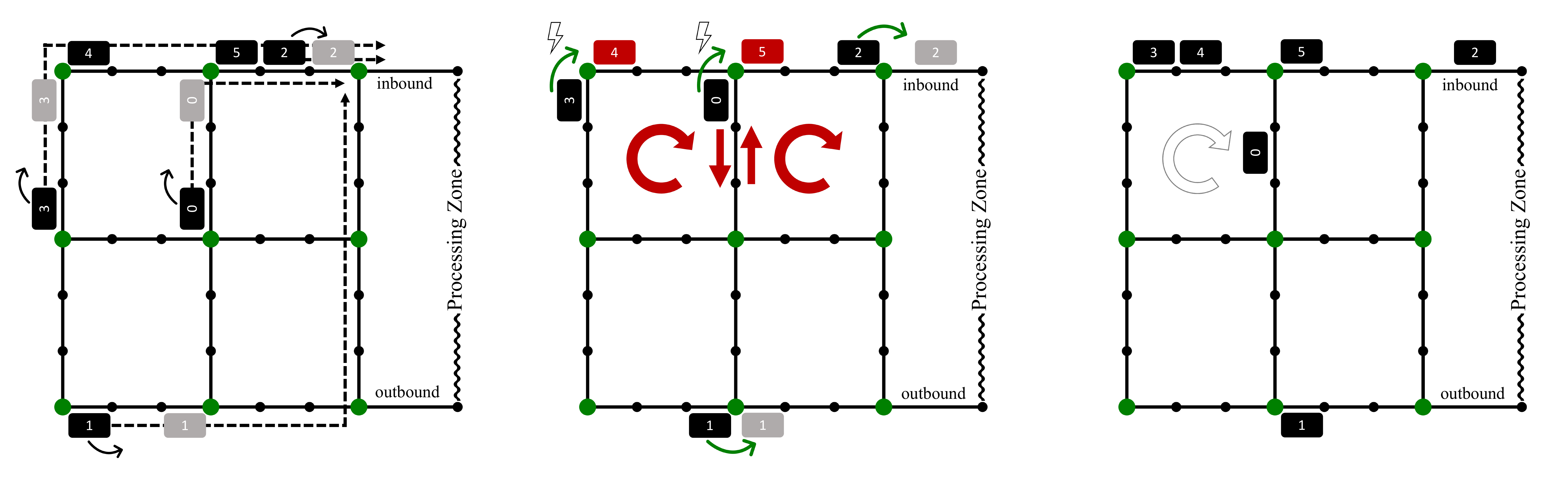}%
    \label{fig:next-move-0}}
    \hfil
    \subfloat[Cycle Construction]{\includegraphics[trim=34cm -1em 36cm 0, clip, width=.32\linewidth]{Figures/circles/circles_path_new.pdf}%
    \label{fig:next-move-1}}
    \hfil
    \subfloat[End of time step]{\includegraphics[trim=68cm -1em 2cm 0, clip, width=.32\linewidth]{Figures/circles/circles_path_new.pdf}%
    \label{fig:next-move-2}}
    \vspace{.5em}
    \caption{Possible movements for one time step}
    \label{fig:next_move}
\end{figure*}

The concept of cycles works in all QCCD architectures, in which closed loops can be formed. In the following, we discuss how to implement a corresponding approach, again, focused on grid-type architectures.
Usually, not all ions are needed in the processing zone at the same time and, thus, we need to prioritize the ions that are needed immediately. To this end, we discuss how to decide which ions are shuttled at the same time and, with that, which cycles are performed.

We start by creating a ranking of the ion chains in the system, which we will call \emph{priority queue}. The priority queue is determined with the help of the \emph{Gate Selection} step explained in \autoref{sec:gate-selection}. As discussed before, the front layer of the Dependency Graph contains all gates that are executable next, which allows us to pick the most favorable next gate. Accordingly, the corresponding ion chains of this gate will be placed at the top of the priority queue. Then, the first gate is disregarded and its node deleted from the Dependency Graph. This leads to a new front layer, from which the next chains in the queue can be determined. If the chain is already present in the priority queue, it is not added again. The priority queue is filled this way until either a maximum number of ion chains is specified or there is no node left in the Dependency Graph.

\vspace{.3em}
\begin{example}
    Revisit the state of the ion chains discussed in \autoref{ex:qft_best_gate} and the corresponding Dependency Graph in \autoref{fig:dag_dep}.
    For simplicity, let all chains $c_0, c_1$ and $c_2$ hold exactly one ion $i_0, i_1$ and $i_2$.
    Since node $6$ was picked as the next gate, ion chain $c_1$ holds the first spot in the priority queue. To find the next chain in the priority queue, we successively remove all nodes which act on chain $c_1$ and only depend on the previous one (these are all colored blue in \autoref{fig:dag_dep}) from the Dependency Graph. This continues until node $16$ which also depends on $c_2$. At this point, depending on which chain is closer to the processing zone, either $c_0$ or $c_2$ is taking the next spot in the priority queue. The other one remains in the last spot. 
\end{example}
\vspace{.3em}

With the priority queue determined, we can start scheduling movement operations for all ion chains.
The goal of this approach is to move as many of the required chains as possible to the processing zone. As a first step, we determine the shortest path of each ion chain in the priority queue.
We then move all chains within two junctions on their path as far as possible.
Afterwards, chains may traverse a junction to reach a neighboring linear region.

\vspace{.3em}
\begin{example}\label{ex:time_step0}
    One time step is illustrated in \autoref{fig:next_move}.
    \mbox{\autoref{fig:next-move-0}} displays the first set of movement operations, which shuttle all chains to the outer edge within their linear region. The resulting configuration is indicated in grey, which is the initial configuration in \autoref{fig:next-move-1}.
\end{example}
\vspace{.3em}

If the next edge of an ion chain is not occupied, the chain can simply be shuttled through the respective junction on its shortest path.
In case the next edge is blocked by another chain, cycles are formed along the shortest path of the shuttling chains and all chains on the cycle are moved in the same direction.

\vspace{.3em}
\begin{example}
    As already discussed in \autoref{ex:time_step0}, the resulting configuration in \autoref{fig:next-move-0}~(grey) provides the initial situation for \autoref{fig:next-move-1}. Four of the six considered chains ($c_0$, $c_1$, $c_2$ and $c_3$) move on their shortest path to the processing zone, as indicated in the illustration. Within one time step, the chains are allowed to move through one junction, if the path is not blocked by another chain. In this case, chains $c_1$ and $c_2$ are free to move over their respective junction, while chains $c_0$ and $c_3$ are both blocked by the additional chains $c_5$ and $c_4$~(red).
\end{example}
\vspace{.3em}

Cycles may overlap and attempt to move chains in opposite directions. To avoid this, we allow only non-overlapping cycles to move at the same time, and, in case of conflicting cycles, prioritize by the position of the chain within the priority queue.

\vspace{.3em}
\begin{example}
     In \autoref{fig:next-move-1}, two cycles are formed along the shortest paths of both $c_0$ and $c_3$, analogous to the description in \autoref{fig:circle}. The two cycles result in a conflict, because they are acting in opposite directions on their shared linear region (indicated by red arrows). The ranking of $c_0$ and $c_3$ in the priority queue then determines, which of the two cycles is performed. Since $c_3$ is needed in the processing zone before $c_0$, the left cycle in \autoref{fig:next-move-1} is prioritized and executed accordingly. The resulting configuration after the completion of one time step is provided in \autoref{fig:next-move-2}.
\end{example}
\vspace{.3em}

However, if all ion chains in the priority queue moved, the chain closest to the processing zone would always be the first one arriving.
To ensure that chains move to the inbound edge in the correct order, the priority queue of the algorithm has to be refined further. Chains only move along their shortest path, and potentially form cycles, if all chains which are needed prior to the considered chain are closer to the inbound edge. In other words, the first chain in the queue always moves, the second only if it is further away from the inbound edge than the first chain and so on. 

As soon as the required ions are present in the processing zone, the gate execution begins. Depending on the specifics of the processing zone, the path towards and into the processing zone may be blocked while the gate is being processed. This can take up to several time steps, again, depending on the specifics of the processing zone and the type of gate.
Once a gate is processed, the corresponding node is removed from the Dependency Graph and the compilation returns to selecting the next best gate. The compiler then constructs an updated Dependency Graph by reevaluating the new positions of ion chains in the system. This continues until the complete quantum circuit is executed on the device.

Cycle-based shuttling provides an efficient heuristic to generate schedules even for larger QCCD architectures.
To validate this claim, the following section evaluates the proposed implementation.

\begin{table*}[t]
    \centering
    \caption{Results of the Empirical Evaluation}
    \label{tab:benchmarks}    
    \sisetup{round-mode=places,round-minimum=0.1}
    \resizebox{\linewidth}{!}{
        \begin{tabular}{@{} c @{} c c @{\hspace{4.5em}} r r r S[table-auto-round,table-format=4.1] @{\hspace{3em}} r r S[table-auto-round,table-format=5.1] @{\hspace{3em}} r r r @{\hspace{3em}} r r r @{}}
            \toprule
            \multicolumn{3}{c}{\textbf{Architecture}} & \multicolumn{4}{c}{\textbf{Full Register Access}} & \multicolumn{3}{c}{\textbf{GHZ}} & \multicolumn{3}{c}{\textbf{Graph}} & \multicolumn{3}{c}{\textbf{QFT}} \\
            \cmidrule(r{2.5em}){1-3}\cmidrule(r{1.5em}){4-7} \cmidrule(r{1.5em}){8-10}\cmidrule(r{1.5em}){11-13}\cmidrule{14-16}
            ~~~~Type~~~~~ & $m$ $n$ $v$ $h$ & {$|C|$/$|E_M|$} & $G$ & {$\hat{T}\textrm{min}$} & {$\hat{T}$} & {$t_\textrm{CPU}$ [s]} 
            & $G$ & {$\hat{T}$} & {$t_\textrm{CPU}$ [s]} 
            & {$G$} & {$\hat{T}$} & {$t_\textrm{CPU}$ [s]} 
            & {$G$} & {$\hat{T}$} & {$t_\textrm{CPU}$ [s]} \\
            \midrule
            
        \multirow{6}{*}{\shortstack[c]{\includegraphics[angle=90,width=.05\textwidth]{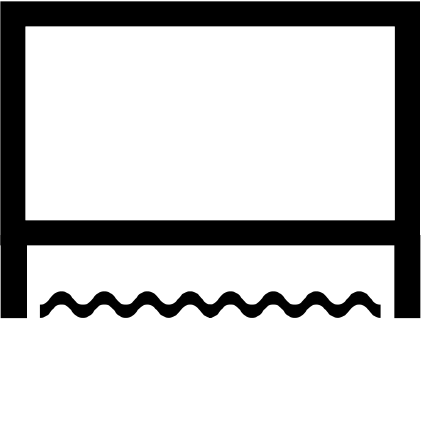} \\Racetrack \vspace{1em}}}
        & 2 2 1 \phantom{1}5 & 6/12  & 6 & 12.5 & 14.6 & 0.4 & 100 & 110.2 & 4.3 & 110 & 122.4 &  4.98 & 250 & 280.2 & 19.5 \\
        & 2 2 1 11 & 12/24  & 12 & 21.9 & 39.5 & 1.8 & 220 & 242.2 & 19.1 & 212 & 236.2 & 19.6 & 1039 & 1178.9 & 740.0 \\
        & 2 2 1 19 & 20/40  & 20 & -- & 95.4 & 6.8 & 380 & 421.5 & 68.8 & 360 & 414.2 & 70.6 & 2901 & 3388.8 & 14502.6 \\
        & 2 2 1 29 & 30/60  & 30 & -- & 185.3 & 18.9 & 538 & 655.8 & 216.4 & 538 & 650.5 & 214.8 & 6775 & 7984.0 & 189710.8\\
        & 2 2 1 39 & 40/80 & 40 & -- & 336.3 & 40.7 & 780 & 914.2 & 540.3 & 720 & 957.5 & 449.8 & 12315 & $^*$70470.4 & $^*$46012.2 \\
        \midrule

        \multirow{6}{*}{\shortstack[c]{\includegraphics[angle=90,width=.05\textwidth]{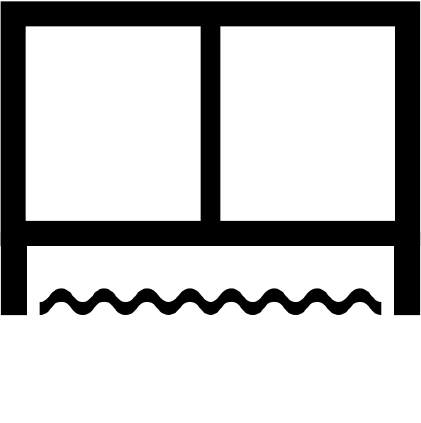}\\Vertical Grate}}
        & 2 4 1 1 & 5/10  & 5 & 11.1 & 12.9 & 0.42 & 80 & 89.7 & 3.2 & 91 & 103.5 & 3.7 & 168 & 321.9 & 10.4 \\
        & 2 6 1 1 & 8/16  & 8 & 14.9 & 18.5 & 0.69 & 140 & 158.0 & 8.4 & 144 & 162.0 & 8.7 & 463 & 526.4 & 244.7 \\
        & 2 8 1 1 & 11/22  & 11 & -- & 28.1 & 1.28 & 200 & 232.1 & 17.2 & 197 & 224.0 & 17.3 & 873 & 995.2 & 1334.9 \\
        & 2 10 1 1 & 14/28  & 14 & -- & 36.7 & 2.05 & 260 & 294.5 & 29.8 & 250 & 280.2 & 30.5 & 1411 & 1624.6 & 5310.6 \\
        & 2 10 5 5 & 70/140  & 70 & -- & 360.1 & 139.8 & 1258 & 1422.2 & 2488.2 & 1258 & 1418.8 & 2490.0 & 29479 & $^*$187761.2 & $^*$82109.4 \\
        \midrule
        
        \multirow{6}{*}{\shortstack[c]{\includegraphics[angle=90,width=.05\textwidth]{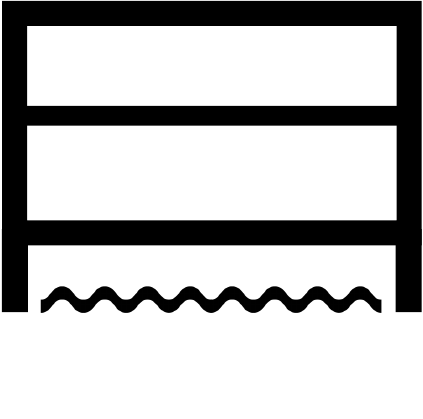}\\Horizontal Grate}}
        & 4 2 1 1 & 5/10  & 5 & 10.7 & 11.4 & 0.30 & 80 & 88.3 & 3.1 & 91 & 103.1 & 3.5 & 168 & 214.5 & 6.6 \\
        & 6 2 1 1 & 8/16  & 8 & 15.7 & 24.4 & 0.78 & 140 & 154.0 & 7.6 & 144 & 160.0 & 8.1 & 463 & 520.4 & 239.5 \\
        & 8 2 1 1 & 11/22  & 11 & -- & 33.7 & 1.44 & 200 & 220.3 & 29.8 & 197 & 220.1 & 15.8 & 873 & 987.2 & 1336.8 \\
        & 10 2 1 1 & 14/28  & 14 & -- & 46.6 & 2.23 & 260 & 286.2 & 27.3 & 250 & 278.9 & 28.0 & 1411 & 1596.6 & 5291.8 \\
        & 10 2 5 5 & 70/140  & 70 & -- & 599.1 & 24.7798 & 1258 & 1448.7 & 2383.2 & 1258 & 1450.5 & 2382.9 & 29479 & $^*$205201.2 & $^*$364680.1 \\
        \midrule

        \multirow{6}{*}{\shortstack[c]{\includegraphics[angle=90,width=.05\textwidth]{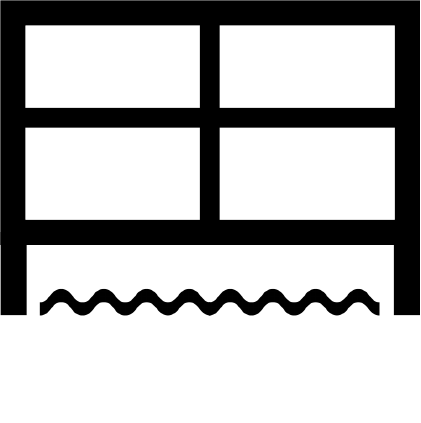}\\\hspace{-.9em}Lattice}}
        & 3 3 1 1 & 6/12  & 6 & 11.0 & 11.5 & 0.42 & 100 & 110.8 & 4.3 & 110 & 122.2 & 4.9 & 250 & 280.2 & 56.1 \\ 
        & 4 4 1 1 & 12/24  & 12 & 20.2 & 26.6 & 1.25 & 220 & 245.6 & 19.4 & 212 & 239.0 & 19.6 & 1039 & 1182.0 & 2192.1 \\
        & 5 5 1 1 & 20/40  & 20 & -- & 49.2 & 3.22 & 380 & 422.0 & 68.0 & 360 & 401.4 & 69.5 & 2901 & 3341.1 & 14802.8 \\
        & 6 6 1 1 & 30/60  & 30 & -- & 62.3 & 6.32 & 580 & 650.2 & 204.6 & 538 & 602.2 & 205.5 & 6775 & 7904.6 & 192553.8\\
        & 10 10 1 1 & 90/180 & 90 & -- & 229.8 & 110.1 & 1780 & 1978.2 & 4821.1 & 1618 & 1815.6 & 4881.1 & 40919 & $^*$102394.2 & $^*$51662.0 \\

        \bottomrule
        \bottomrule
        \noalign{\smallskip}
        \multicolumn{16}{r@{}}{\textsuperscript{*}Benchmark without \emph{Gate Selection} step}
        \end{tabular}
    }
\end{table*}

\section{Empirical Evaluation}
\label{sec:evaluation}
This section provides the results of an evaluation of the proposed approach. The methods described above have been implemented as part of the open-source \emph{Munich Quantum Toolkit} at \mbox{\url{https://github.com/cda-tum/mqt-ion-shuttler}}.
In our evaluations, we considered different architectures following a grid structure. On each architecture, we generated shuttling schedules for four different circuits, each compiled to real hardware implementations of state-of-the-art trapped-ion quantum computers.
In line with the previous sections, we first discuss the specific realization of individual steps in the compilation flow. This is followed by the description of the detailed setup that was used in the evaluation as well as the obtained results.

\subsection{Realization of Compilation Steps}
After the removal of all SWAP gates, the quantum circuits were translated to a common native gate set consisting of RZZ, RZ, RY, and RX. This particular gate set also finds application in real-world quantum computers, e.g., in QCCD devices offered by Quantinuum~\cite{moses2023race}. The considered \mbox{high-level} circuits are obtained via MQT Bench~\cite{Quetschlich_2023}, which are translated and compiled using existing functionality from the Python module pytket~\cite{Sivarajah_2020}. To realize the circuit optimization step, the pytket module also provides optimization routines to further transpile and reduce the number of gates of each circuit.
The Dependency Graph used in the Gate Selection step is constructed with qiskit~\cite{javadiabhari2024quantum} and uses the corresponding commutation rules.\\
In this evaluation, we only consider a single ion in each ion chain and employ a direct mapping where qubit $q_j$ corresponds to ion $i_j$. This allows us to provide a clear and direct assessment of the shuttling approach, without the complexities introduced by multiple ions per chain. The proposed mapping of $q_j\xrightarrow{}i_j$ is also chosen for its simplicity and to ensure the empirical evaluation actually evaluates the shuttling approach. Depending on the capabilities of the underlying hardware and the considered quantum algorithm, other mappings may provide a benefit.

\subsection{Setup}
The important properties of the grid-like graphs are described by four values $m,n,v,h$: the grid is of size $m \times n$, with $m$ nodes vertically and $n$ nodes horizontally. For the architecture, this means at most~$v$~($h$)~ion~chains can be trapped vertically (horizontally) between to junctions. Accordingly, the graph in \autoref{fig:graph} is denoted as $\{3, 3, 3, 3\}$.
The grid is further extended by two additional edges which represent one outbound edge to a processing zone and one inbound edge leading back to the memory grid.
Using a random starting configuration of ion chains on these grids, we then used the proposed approach to determine the number of time steps~$\hat{T}$ that are sufficient to realize a given quantum circuit, i.e.,~shuttling schedule.

For the evaluation, we consider four access patterns and four different types of architectures.
The access patterns are (with each chain containing exactly one ion)
\begin{itemize}
    \item \enquote{Full register access} where each ion chain is shuttled to the processing zone once,
    \item \enquote{GHZ} where the ion chains are scheduled according to a circuit that realizes the Greenberger–Horne–Zeilinger state for all qubits,
    \item \enquote{Graph} where the ion chains are scheduled according to a circuit that realizes the Graph state for all qubits, and
    \item \enquote{QFT} where the ion chains are scheduled according to the quantum Fourier transform discussed before.
\end{itemize}
The different architecture types are described as follows:
\begin{itemize}
    \item \enquote{Racetrack} which consists of one big ring in which the ion chains can move,
    \item \enquote{Horizontal Grate} where there are only trap sites in parallel to the processing zone except for the bordering linear regions,
    \item \enquote{Vertical Grate} where there are only trap sites perpendicular to the processing zone except the bordering linear regions, and
    \item \enquote{Grid} which has has both parallel and perpendicular trap sites inside.
\end{itemize}

All evaluations were conducted on a machine with an Intel(R) Xeon(R) W-1370P CPU (running at ~\SI{3.22}{\giga\hertz}) and \SI{32}{\gibi\byte} main memory running Python 3.8.10 and averaged over five runs with different random initial configurations.

\subsection{Results}

\autoref{tab:benchmarks} summarizes the results of the evaluation.
The first group of columns describes the architecture, providing the general type, exact dimensions, and the fraction \mbox{$|C|$/$|E_M|$} of occupied edges (sites) to the total number of edges (sites) in the memory zone. For every experiment, exactly half of the edges were occupied by ion chains.
The following four groups of columns present the results for the benchmarks executing \enquote{Full register access}, the \enquote{GHZ} state, the \enquote{Graph} state, and the quantum Fourier transform \enquote{QFT}.
Here, $G$ is the number of gates, $\hat{T}$ is the number of time steps required for the generated shuttling schedule, and $t_\textrm{CPU}$ is the time taken to generate these shuttling schedules.
For small instances of \enquote{Full register access}, we also list the minimal number of time steps $\hat{T}_\textrm{min}$.
The largest instances of the \enquote{QFT} benchmark were conducted without the \enquote{Gate Selection} step to allow the evaluation of all circuits within a maximum time of $200\,000\si{\s}$. Excluding this step reduces the computation time at the expense of more time steps, i.e., a less efficient shuttling schedule.
The results show that the proposed approach is capable of generating shuttling schedules for a variety of \mbox{trapped-ion} hardware. Even for large architectures, the implementation is able to produce efficient shuttling schedules.
Compared to the minimal results obtained by exhaustive search, the proposed heuristic approach performs reasonably well.
Even though the complexity increases for larger architectures, we do not expect the gap between the minimal and heuristic solutions to become relatively larger, as larger architectures offer more potential for finding cycles. Furthermore, increasing the number of ions per ion chain may reduce the amount of necessary shuttling, thus making it possible to support larger architectures and more complex quantum circuits.

\section{Conclusion}
\label{sec:conclusion}

With the QCCD architecture, trapped-ion quantum computers provide a modular design that promises good scalability.
Still, efficient classical design tools are required to tap into this potential.
In this paper, we proposed such a tool for generating efficient shuttling schedules.
To this end, we discussed how to compile a given high-level quantum circuit to the specifics of a QCCD device and, then, use a \mbox{graph-based} abstraction of the underlying hardware to discretize the problem of moving ion chains.
Further, we exploit the topology for conflict-free shuttling through cycles in the graph, combined with choosing the most favorable next gate after each gate execution.
The empirical evaluation confirms that the proposed approach is able to generate efficient shuttling schedules, even for large systems.
The corresponding implementation is freely available as part of the open-source \emph{Munich Quantum Toolkit} at \mbox{\url{https://github.com/cda-tum/mqt-ion-shuttler}} under the MIT license.
Possible future work includes determining more sophisticated initial qubit-ion-mappings and the extension to multiple processing zones.

\section*{\sc Acknowledgments}
This work was funded under the European Union's Horizon 2020 research and innovation programme (DA QC, grant agreement \mbox{No. 101001318} and MILLENION, grant agreement No. 101114305), the State of Upper Austria in the frame of the COMET program, the QuantumReady project within Quantum Austria (managed by the FFG), and was part of the Munich Quantum Valley, which is supported by the Bavarian state government with funds from the Hightech Agenda Bayern Plus.

%%%%%%%%%%%%%%%%%%%%%%%%%%%%%%%%%%%%%%%%%%%%%%%%%%%%%%%%%%%%%%%%%%%%%%%

\printbibliography 

\newpage
\begin{IEEEbiography}[{\includegraphics[width=1in,height=1.25in,clip,keepaspectratio]{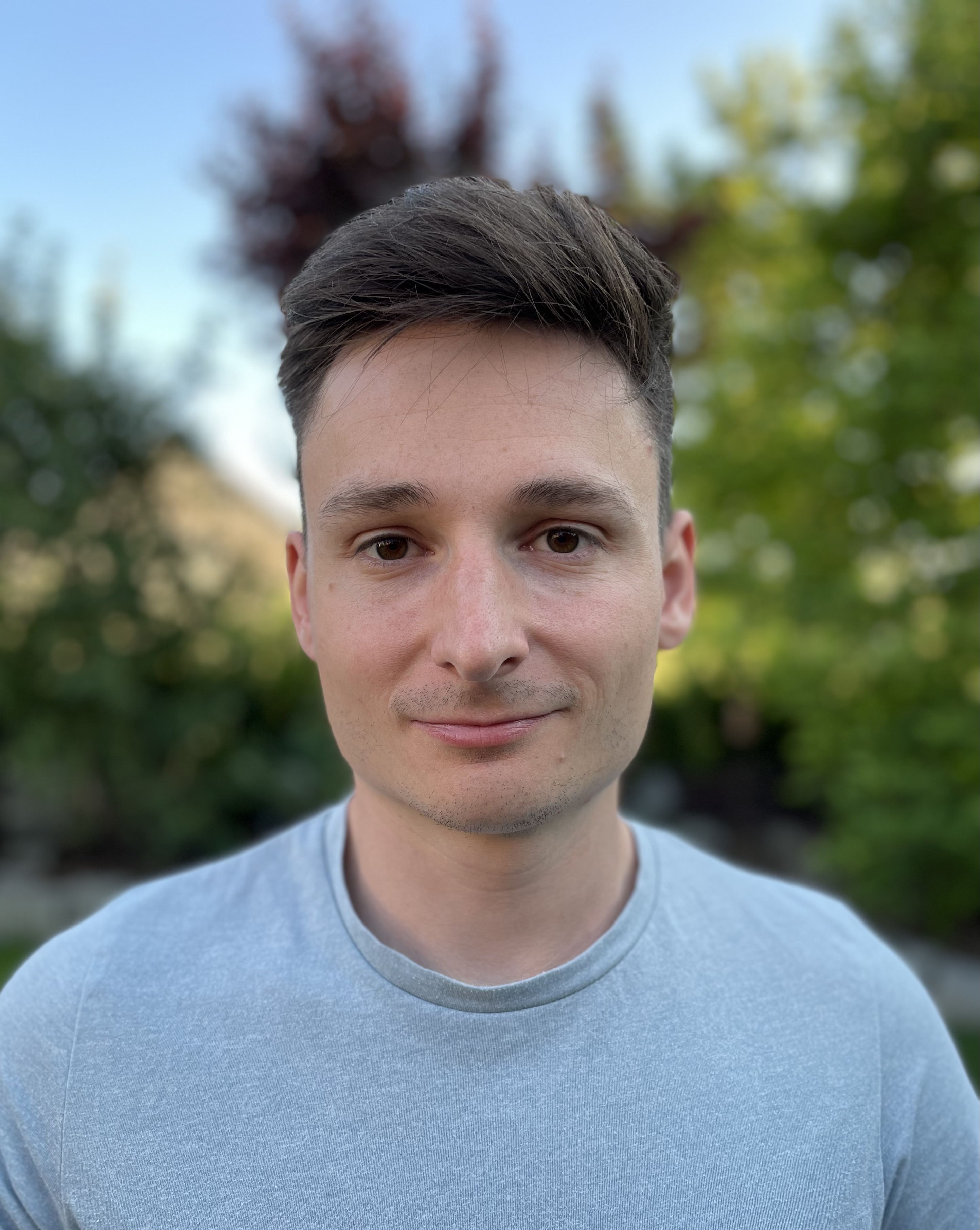}}]{Daniel Schoenberger} received his Master's degree in Physics from the University of Regensburg, Germany, in 2021.
Since 2022, he is a researcher at the Technical University of Munich, Germany, and is working towards his PhD. His research focuses on simulation and compilation for \mbox{trapped-ion} quantum computers.
\end{IEEEbiography}

\begin{IEEEbiography}
[{\includegraphics[width=1in,height=1.25in,clip,keepaspectratio]{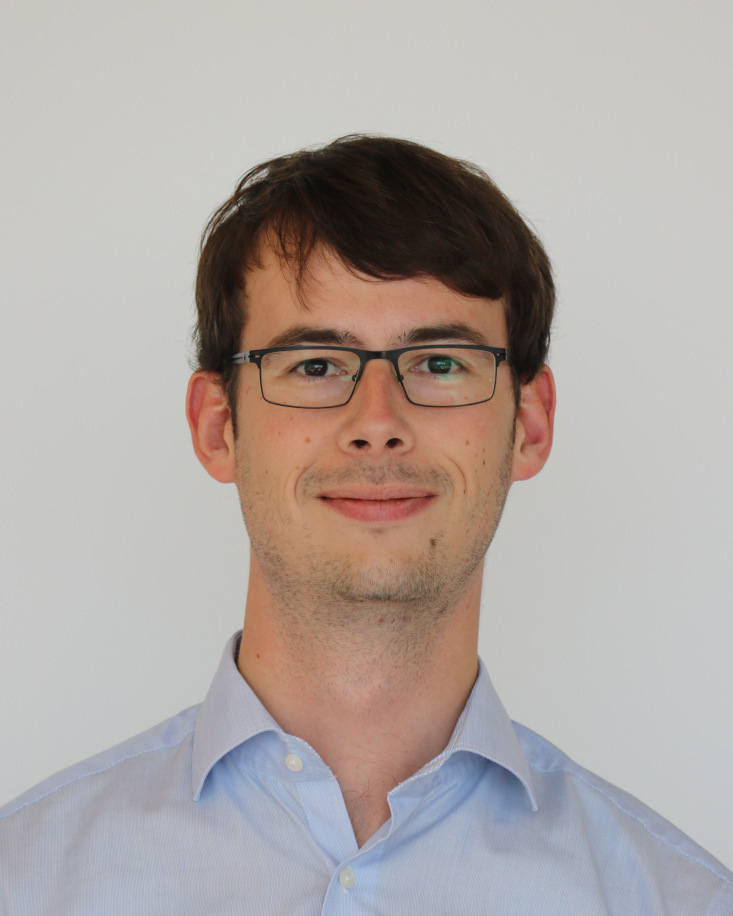}}]{Stefan Hillmich} is a senior researcher for Software Science with a focus on quantum computing and design automation working at the Software Competence Center Hagenberg (SCCH) GmbH since 2023.
He received his B.Sc. and M.Sc from the University of Bremen, Germany, in 2015 and 2018, respectively. In 2022, he received his Dr. techn. degree from the Johannes Kepler University Linz in Austria, where he also worked as PostDoc.
In the areas of quantum computing and design automation, he published several papers in international venues such as TCAD, TQC, DAC, ASP-DAC, DATE, and ICCAD.
\end{IEEEbiography}

\begin{IEEEbiography}
[{\includegraphics[width=1in,height=1.25in,clip,keepaspectratio]{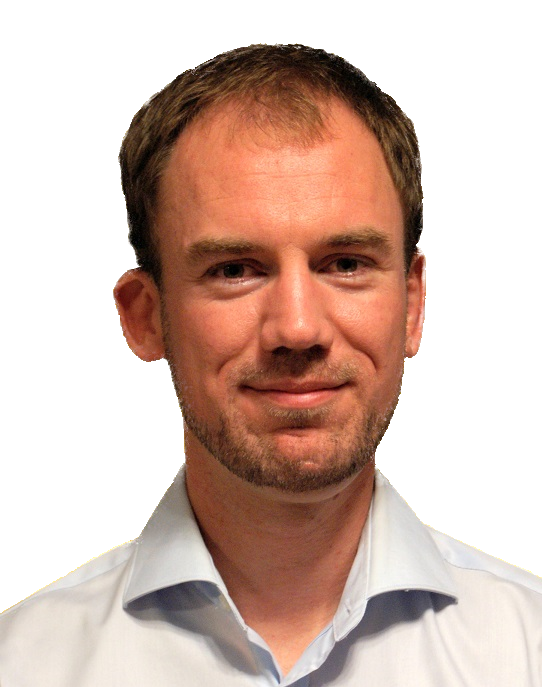}}]{Matthias Brandl} received his BSc in electrical engineering in 2008, his MSc in microelectronics in 2010, and his Diploma in technical physics in 2011 all from Technical University of Vienna, Austria. In 2017, he received his PhD in experimental quantum physics on the topic of cryogenic trapped ion quantum computing from the University of Innsbruck, Austria. In 2017, he joined Infineon Technologies Austria AG in Villach, Austria and since 2018, he is working at Infineon Technologies AG in Munich, Germany. Until 2022, his focus was MMICs. He worked on the application side as well as on the concept side of radar sensors. Since 2022, he is the system architect for trapped-ion quantum computing and his focus is on electronics development and control of large ion traps.
\end{IEEEbiography}

\begin{IEEEbiography}
[{\includegraphics[width=1in,height=1.25in,clip,keepaspectratio]{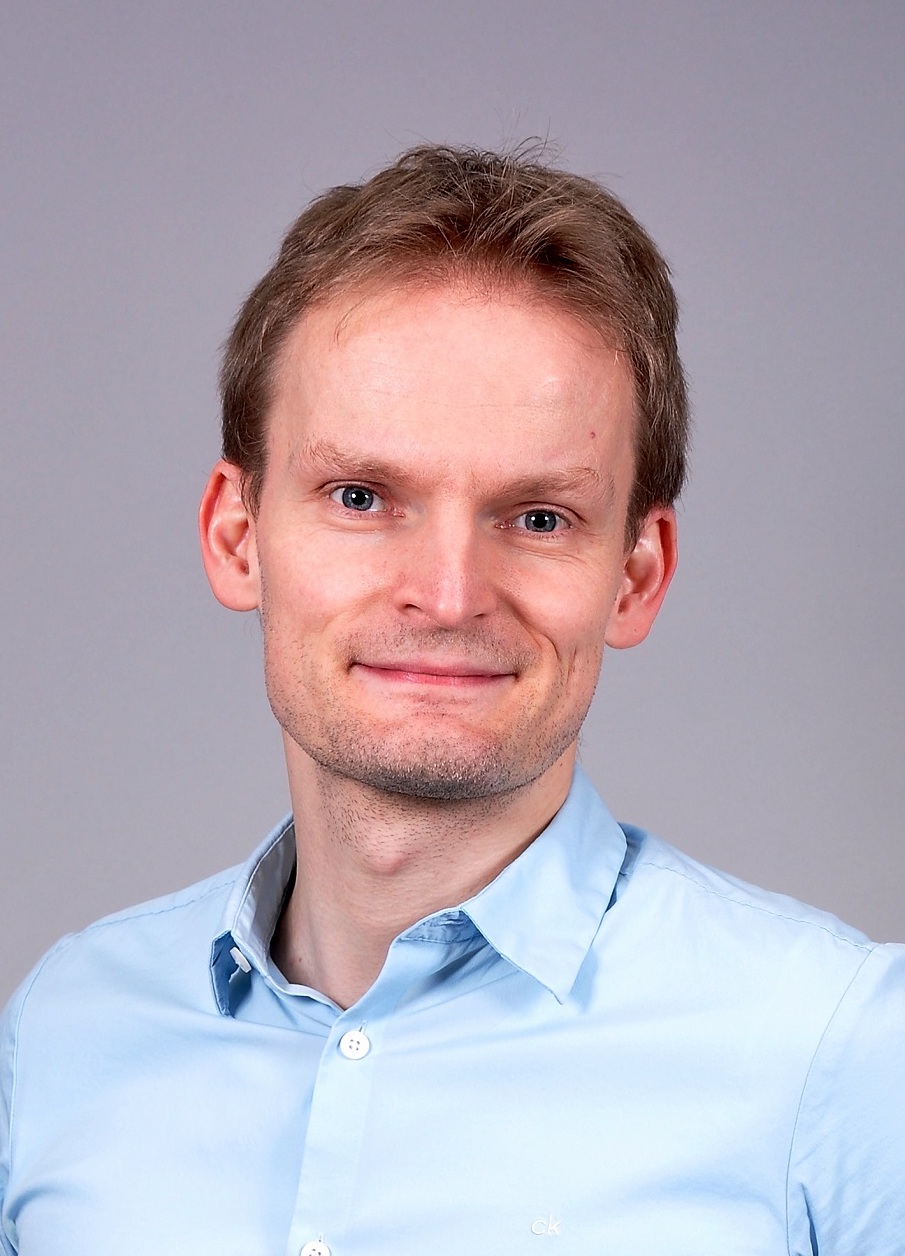}}]{Robert Wille} is a Full and Distinguished Professor at the Technical University of Munich, Germany, and Chief Scientific Officer at the Software Competence Center Hagenberg, Austria. He received the Diploma and Dr.-Ing. degrees in Computer Science from the University of Bremen, Germany, in 2006 and 2009, respectively. Since then, he worked at the University of Bremen, the German Research Center for Artificial Intelligence (DFKI), the University of Applied Science of Bremen, the University of Potsdam, and the Technical University Dresden. From 2015 until 2022, he was Full Professor at the Johannes Kepler University Linz, Austria, until he moved to Munich. His research interests are in the design of circuits and systems for both conventional and emerging technologies. In these areas, he published more than 400 papers and served in editorial boards as well as program committees of numerous journals/conferences such as TCAD, \mbox{ASP-DAC}, DAC, DATE, and ICCAD. For his research, he was awarded, e.g., with Best Paper Awards, e.g., at TCAD and ICCAD, an ERC Consolidator Grant, a Distinguished and a Lighthouse Professor appointment, a Google Research Award, and more.
\end{IEEEbiography}

\end{document}